\documentclass[3p]{elsarticle}

\usepackage{lineno,hyperref}
\usepackage{caption}
\usepackage{subcaption}

\usepackage{amsmath}
\usepackage{multirow}
\usepackage{graphicx}
\usepackage{float}
\usepackage{xcolor}
\usepackage[linesnumbered,ruled,vlined]{algorithm2e}
\usepackage{adjustbox}
\RestyleAlgo{ruled}
\SetKwComment{Comment}{/* }{ */}
\usepackage{tabularx}
\usepackage{rotating}

\usepackage[nolist]{acronym}
\begin{acronym}[AoI]
\acro{AoI}{Area of Interest}
\end{acronym}
\begin{acronym}[CAM]
\acro{CAM}{Cooperative Awareness Message}
\end{acronym}
\begin{acronym}[CBR]
\acro{CBR}{Channel Busy Rate}
\end{acronym}
\begin{acronym}[ETSI]
\acro{ETSI}{European Telecommunications Standards Institute}
\end{acronym}
\begin{acronym}[ITS]
\acro{ITS}{Intelligent Transport Systems}
\end{acronym}
\begin{acronym}[DCC]
\acro{DCC}{Decentralized Congestion Control}
\end{acronym}
\begin{acronym}[CA]
\acro{CA}{Cooperative Awareness}
\end{acronym}
\begin{acronym}[CAM]
\acro{CAM}{Cooperative Awareness Message}
\end{acronym}
\begin{acronym}[CBF]
\acro{CBF}{Contention-based Forwarding}
\end{acronym}
\begin{acronym}[DEN]
\acro{DEN}{Decentralized Environmental Notification}
\end{acronym}
\begin{acronym}[DENM]
\acro{DENM}{Decentralized Environmental Notification Message}
\end{acronym}
\begin{acronym}[DPD]
\acro{DPD}{Duplicate Packet Detection}
\end{acronym}
\begin{acronym}[DPL]
\acro{DPL}{Duplicate Packet List}
\end{acronym}
\begin{acronym}[GPC]
\acro{GPC}{Geographically-aware CBF Packet Cancellation}
\end{acronym}
\begin{acronym}[FCD]
\acro{FCD}{Floating Car Data}
\end{acronym}
\begin{acronym}[FoT]
\acro{FoT}{Forward on Time}
\end{acronym}
\begin{acronym}[FoT+]
\acro{FoT+}{Forward on Time+}
\end{acronym}
\begin{acronym}[PDR]
\acro{PDR}{Packet Delivery Ratio}
\end{acronym}
\begin{acronym}[RSU]
\acro{RSU}{Road-side Unit}
\end{acronym}
\begin{acronym}[RHW]
\acro{RHW}{Road Hazard Warning}
\end{acronym}
\begin{acronym}[S-CBF]
\acro{S-CBF}{Slotted CBF}
\end{acronym}
\begin{acronym}[SHB]
\acro{SHB}{Single-Hop Broadcasting}
\end{acronym}
\begin{acronym}[TRC]
\acro{TRC}{Transmit Rate Control}
\end{acronym}
\begin{acronym}[VANET]
\acro{VANET}{Vehicular ad hoc Network}
\end{acronym}

\journal{Computer Networks}

\begin{document}

\begin{frontmatter}

\title{Studying and improving the performance of ETSI ITS Contention-Based Forwarding (CBF) in urban and highway scenarios: S-FoT+}

\tnotetext[mytitlenote3]{\textbf{Published as: Oscar Amador; Ignacio Soto; Maria Calderon; Manuel Urueña. Studying and improving the performance of ETSI ITS contention-based forwarding (CBF) in urban and highway scenarios: S-FoT+. Computer Networks 233-109899, pp.1-17, 2023.  The final version of record is available at \href{https://doi.org/10.1016/j.comnet.2023.109899}{https://doi.org/10.1016/j.comnet.2023.109899}}}
\tnotetext[mytitlenote]{This work was partially supported by the Agencia Estatal de Investigaci\'on (AEI, Spain) through the ACHILLES project (PID2019-104207RB-I00/AEI/10.13039/501100011033)}
\tnotetext[titlenote2]{***We gratefully acknowledge support from the Swedish Knowledge Foundation (KKS) "Safety of Connected Intelligent Vehicles in Smart Cities -- SafeSmart" project (2019--2024), the Swedish Innovation Agency (VINNOVA) "Emergency Vehicle Traffic Light Preemption in Cities -- EPIC, ref. No. 2020-02945" (2020--2022), and the ELLIIT Strategic Research Network.}


\author[Halmstad]{Oscar Amador}
\ead{oscar.molina@hh.se}

\author[UPM]{Ignacio Soto\corref{mycorrespondingauthor}}
\cortext[mycorrespondingauthor]{Corresponding author}
\ead{ignacio.soto@upm.es}

\author[UC3M]{Maria Calderon}
\ead{maria@it.uc3m.es}

\author[UNIR]{Manuel Urue\~na}
\ead{manuel.uruena@unir.net}

\address[Halmstad]{School of Information Technology; Halmstad University; Halmstad 30118; Sweden}
\address[UPM]{Departamento de Ingenier\'{\i}a de Sistemas Telem\'aticos; Universidad Polit\'ecnica de Madrid; 28040 Madrid (Madrid); Spain}
\address[UC3M]{Departamento de Ingenier\'{\i}a Telem\'atica; Universidad Carlos III de Madrid; 28911 Legan\'es (Madrid); Spain}
\address[UNIR]{Escuela Superior de Ingenieros y Tecnolog\'{\i}a, Universidad Internacional de la Rioja; 26006 Logro\~no; Spain}

\begin{abstract}
This paper evaluates the performance of ETSI ITS Contention-Based Forwarding (CBF) and ETSI Simple GeoBroadcast forwarding while disseminating warning messages over a Geographical Area in highway and urban scenarios. Our experimental evaluation considers the complete ETSI ITS architecture including the Decentralized Congestion Control (DCC) mechanism. We propose \textcolor{black}{an enhanced CBF mechanism, named \mbox{S-FOT+}, which combines several improvements to the ETSI CBF algorithm. S-FoT+ has a} similar or better performance than the ETSI forwarding algorithms regarding both reliability and end-to-end delay while requiring much fewer transmissions. The improvements are equally effective and efficient in both urban and highway scenarios with large Destination Areas. Finally, we evaluate the trade-offs that stem from using multi-hop broadcast mechanisms in urban settings with smaller Destination Areas when compared to single-hop broadcast. Results show that multi-hop mechanisms significantly improve coverage at the cost of an increased number of transmissions.  
\end{abstract}


\end{frontmatter}


\section{Introduction}
\label{sec:intro}
\acp{VANET} enable communication among vehicles, and they are considered a basic building block for the development of \acf{ITS}. Vehicles share information among themselves and with the infrastructure to improve driving safety and traffic efficiency, as well as to provide comfort and infotainment applications to passengers.

Driving safety is a relevant cooperative service where vehicles timely disseminate warning messages over a Geographical Area (GeoArea) to prevent accidents. Examples of emerging \ac{ITS} safety applications are precrash warning, lane change warning, road obstacle detection, road status, adaptive traffic lights, or accident notifications. In Europe, \ac{ITS} specifications are being standardized by \ac{ETSI} with support from industry and academia. 

\ac{RHW} \cite{etsiRHS} is an event-based road message dissemination service based on \acp{DENM} to provide relevant alert information to the affected road users inside a GeoArea. \acp{DENM} are distributed using the ETSI GeoNetworking services \cite{etsiNewGeoNetworking}. In particular, multi-hop broadcast (i.e., GeoBroadcast) is seen as a good scheme for event-driven data dissemination over a GeoArea (i.e., retransmitting these messages until all vehicles in the Destination Area receive them). The ETSI GeoBroadcast protocol offers several forwarding algorithms that control how messages are retransmitted. For the case of message dissemination within the Destination Area (i.e., area forwarding) ETSI has specified two basic forwarding strategies: Simple GeoBroadcast forwarding and \ac{CBF}. The specification also includes several non-area forwarding strategies that have not been considered in this work. 

The ETSI Simple GeoBroadcast Forwarding algorithm is a simple flooding mechanism, where all \textcolor{black}{vehicles} rebroadcast a packet as soon as they receive it. Besides, each forwarder employs a \ac{DPD} strategy to track packets that have been previously rebroadcast.  The \ac{DPD} strategy is based on the source \textcolor{black}{vehicle} address, the sequence number of the packet, a maximum hop limit, and a maximum lifetime. At most, each vehicle in the area forwards the packet once.

ETSI \ac{CBF} is a timer-based forwarding strategy. Once a vehicle receives a multi-hop packet targeted to a Destination Area, it does not immediately forward the received packet but stores it in a dedicated packet buffer (namely, the CBF buffer). Each queued packet has an associated timer. On the expiry of its timer, the associated packet is rebroadcast. The timer duration is scaled inversely proportional to the distance between the previous sender and the current vehicle. On the contrary, the packet is removed from the \ac{CBF} buffer if another vehicle rebroadcasts the packet before the timer expires. As a result, vehicles further away tend to retransmit a packet earlier, while nearby vehicles cancel the buffered packet according to the \ac{CBF} overhearing.

The ETSI ITS specifications define a complete architecture where the ETSI GeoNetworking protocol~\cite{etsiNewGeoNetworking} is located at the Networking\&Transport layer. The Facilities layer relies on the services provided by the GeoNetworking protocol to offer its services to ITS applications. At the Facilities layer, \ac{CA}~\cite{etsiCA} and Decentralized Environmental Notification (DEN)~\cite{etsiDEN} services are essential building blocks for \ac{ITS} applications. The Cooperative Awareness service uses \acp{CAM} which are periodically broadcast to one-hop neighbors, conveying status information (e.g., position and heading of the ego-vehicle). In addition, \acp{DENM} enable DEN services.  A DENM is triggered if an exceptional traffic condition or road hazard is recognized, and the message contains information on the nature of the hazard and its location~\cite{etsiRHS}.

The ETSI GeoNetworking protocol can work over different access technologies in the dedicated 5.9~GHz band (Access Layer), with IEEE 802.11p/ETSI ITS-G5 being one possible technology~\cite{etsiMediaDependentG5}. The Access Layer includes a \ac{DCC} mechanism to control channel load and guarantee that the radio medium works at an efficient regime. The latest version of the ETSI specification~\cite{etsiNewDcc} introduces an adaptive variant of \ac{DCC} based on the LIMERIC algorithm~\cite{Limeric2013}. This variant uses a \acf{TRC} strategy to independently adjust the packet sending rate at each transmitter. To this end, a linear control system is employed using channel occupancy as input. \ac{DCC} at the access layer measures the \ac{CBR}, i.e., the time percentage that the channel is busy, which is a metric of channel occupancy. The \ac{DCC} mechanism is a gatekeeper that schedules packets stored in the \ac{DCC} queues waiting to be dequeued according to the allowed sending rate. There is also a \ac{DCC} component at the Facilities Layer that, together with the \ac{CA} service, modulates \ac{CAM} frequency according to vehicle dynamics and channel occupancy level.  The \ac{DCC} components at different layers communicate via the DCC cross-layer Management Plane. 

The use of CBF as a multi-hop GeoBroadcast protocol poses several challenges that are still open. Good coverage over the whole Destination Area is needed to adequately alert all affected vehicles while maintaining a reasonable end-to-end delay and controlling the overhead introduced by the dissemination protocol. Furthermore, the behavior of the \ac{CBF} protocol has to be analyzed not as an isolated protocol, but as a component of the whole ETSI ITS architecture. For instance, the interaction with \ac{DCC} may impair the performance of \ac{CBF}. Additionally, the \ac{CBF} protocol must show adequate performance not only on highways but also in urban scenarios.

Finally, we explore the possibility of using multi-hop GeoBroadcast to improve coverage at short distances or smaller GeoAreas for specific applications. Works in the literature~\cite{LimericSta,OurAccess} show that \ac{SHB} messages can reach short distances with high success rates in highway environments at various vehicle densities. These works show a proportionally inverse relation between distance and success rates, with values in the range of 60-70\% at around 300\,m. These rates can be lower in urban environments, where obstacles and propagation phenomena hinder the reliability of single-hop schemes. In these scenarios, multi-hop GeoBroadcast could become a viable option to disseminate information successfully if an application requires to service Destination Areas that are expected to be covered by single-hop mechanisms in more favorable circumstances.

This article extends our previous work~\cite{Amador2022} by analyzing the behavior of the ETSI ITS CBF and Simple GeoBroadcast protocols in urban scenarios (with simulations in central Madrid). We also introduce two new enhancements to the \ac{ETSI} \ac{ITS} \ac{CBF} protocol, and evaluate them in both highway and urban scenarios. Therefore, the contributions of this work can be summarized as follows:
\begin{enumerate}
    \item \textcolor{black}{\textit{\ac{ETSI} CBF in urban scenarios:}} Evaluation of \ac{ETSI} \ac{CBF} and proposed improvements in urban scenarios to compare with the performance of those protocols in highway scenarios identified in our previous work~\cite{Amador2022}. 
    \item \textit{FoT+:} A further improvement to our \ac{FoT} mechanism to maximize the time packets spent in the CBF buffer, where they can be cancelled, before being transmitted. 
    \item \textit{Slotted CBF  (S-CBF):} Another improvement to the \ac{ETSI} \ac{CBF} protocol to avoid collisions caused by the \ac{CBF} timer value distribution at long ranges. \textcolor {black} {Slotted CBF mechanism built on top of FoT is referred as S-FOT and on top of FoT+ as S-FOT+. Therefore, S-FoT+ is the name of the updated CBF mechanism including all our proposals.}
    \item \textcolor{black}{\textit{\ac{ETSI} Simple Geobroadcast:}} Evaluation of \ac{ETSI} Simple GeoBroadcast forwarding algorithm in \textcolor{black}{both} highway and urban scenarios.
    \item \textcolor{black}{\textit{Single-hop versus multi-hop broadcast in small Destination Areas}: }A comparison of performance between multi-hop broadcast mechanisms and single-hop broadcast to cover small Destination Areas in urban scenarios.
\end{enumerate}

The remainder of this article is organized as follows: Section \ref{sec:soa} reviews the \ac{ETSI} \ac{CBF} improvements proposed in the literature, summarizing also the proposals from our previous work~\cite{Amador2022}. Section \ref{sec:improvements} proposes two further improvements to \ac{CBF}: Slotted CBF and FoT+. Section~\ref{sec:highway} evaluates these additional proposals in highway scenarios and compares their results with ETSI \ac{SHB}, ETSI CBF, ETSI Simple GeoBroadcast forwarding, and our previous improvements~\cite{Amador2022}. Section~\ref{sec:urban} analyzes the performance of ETSI CBF, ETSI Simple GeoBroadcast and all the proposed improvements in urban scenarios, and compares the performance of multi-hop broadcast mechanisms to \ac{ETSI} \ac{SHB}~\cite{etsiNewGeoNetworking} in terms of coverage in a small area. Finally, Section~\ref{sec:conclusions-fw} summarizes the main results of this paper.

\section{State of the art and background}
\label{sec:soa}

\subsection{Multi-hop dissemination in vehicular networks}

Multi-hop dissemination in vehicular networks has been extensively discussed in recent years. Recent surveys \cite{Shahwani2022, Sanguesa:2016} present a systematic review of proposed techniques and their application to both safety and infotainment applications. 

Two main families of multi-hop dissemination approaches can be identified, namely sender-based and receiver-based. In sender-based protocols, the source (first hop) or the current forwarder (following hops) selects a set of next forwarders~\cite{Torrent2007, Sahoo:2011, bai:2009, yoo:2015, Oliveira2017} based on the knowledge they have of their 1-hop neighbors, obtained through \acp{CAM}. The main limitation of these solutions is their sensibility to the fast-changing topology due to the mobility of vehicles. Explicitly selected relays may not receive the message, resulting in poor dissemination of the packet in the Destination Area. On the contrary, in receiver-based approaches, the own receiver decides whether it is included in the set of next forwarders or not. This decision can be probabilistic~\cite{zeng:2018, Abbasi:2020, REINA:2015, Srivastava2020}. Even though probabilistic approaches can exhibit good delay dissemination, they may incur in collisions and inefficient use of the channel if multiple receivers decide to forward the message. To overcome this limitation, the contention-based (also known as delay-based or timer-based) approach seems a good alternative where the receivers contend for becoming a forwarder~\cite{DDT:2000, chuang:2013, lenardi:2007, Baiocchi:2016, Baiocchi2016b}. The contention time (i.e., forwarding delay) can be computed according to different parameters (e.g., distance from the previous sender, vehicle speed, density). The broadcast address is used in contention-based protocols as the destination link address, so once the first \textcolor{black}{vehicle} forwards the message (the contender with the lowest timer) the rest of the contenders inhibit their scheduled transmissions. This indiscriminate election of the next-forwarder candidates may lead to situations where the best forwarders are inhibited \cite{Hajjej2022, Turcanu2020} giving as a result poor dissemination coverage. 

\acf{CBF} \cite{CBF:2003} is a timer-based protocol aiming at spreading a message in a given Destination Area, accepted by the scientific and industrial communities \cite{Kuhlmorgen:2015}, where the contention timer calculation is based on the distance from the candidate forwarder to the previous sender (i.e., the source or the previous forwarder) while keeping the message in the \ac{CBF} buffer, so it can be cancelled if a better forwarder retransmits. \ac{ETSI} \ac{ITS} standardization~\cite{etsiNewGeoNetworking} has specified \ac{CBF} as the default option for multi-hop GeoBroadcast dissemination if the potential forwarder is inside the Destination Area.

\subsection{Background}
\label{sec:previous}

In \cite{Amador2022} we proposed several mechanisms to improve the performance of the \ac{ETSI} \ac{ITS} \ac{CBF} algorithm in highway scenarios. A brief description of the mechanisms, categorized by their functionality, follows (please, refer to \cite{Amador2022} for further details):

\begin{itemize}
\item \textbf{Adding \acf{DPD} to ETSI ITS \ac{CBF}}: This mechanism reduces unnecessary retransmissions caused by duplicated packets. Within \ac{CBF}, DPD functions as follows: when a packet is first received at a vehicle inside the Destination Area, the packet is sent to the upper layers and stored in the \ac{CBF} buffer as a candidate for future retransmission. If copies of the packet are received because another vehicle has retransmitted it again, the copies are discarded (DPD function), and if the original packet is still in the \ac{CBF} buffer, the packet in the \ac{CBF} buffer is also discarded and its scheduled retransmission is cancelled. The ETSI ITS specification~\cite{etsiNewGeoNetworking} explicitly states that \ac{DPD} should not be used with \ac{CBF}, instead the algorithm solely relies on the \ac{CBF} mechanism itself to avoid duplicate retransmissions \textcolor{black}{(i.e., forwarding inhibition by cancelling duplicate packets in the CBF buffer).} \textcolor{black}{The forwarding inhibition of CBF has no memory beyond the time a packet leaves the CBF buffer. This can cause the packet to be received and transmitted multiple times by a vehicle, for example if a neighboring vehicle did not receive the cancellation packet and sends a spurious retransmission. \ac{DPD} provides long-term memory and prevents a packet from being sent more than once by the same vehicle. As} we experimentally found in \cite{Amador2022}, in real conditions the addition of the proposed \ac{DPD} mechanism to CBF greatly reduces retransmissions, in many cases achieving an order of magnitude reduction.     

\item \textbf{Suppression of Greedy Forwarding collisions at the area border}: Through this mechanism, vehicles using the Greedy algorithm as their non-area forwarding algorithm\,\footnote{\textcolor{black}{The ETSI GeoBroadcast protocol considers the possibility of using different forwarding schemes inside the Destination Area (the forwarding vehicle is inside the area), or outside the Destination Area (the forwarding vehicle is outside the area and forwards a message towards the area from a source on the outside). The default schemes are CBF for area forwarding, and Greedy for non-area forwarding.}} do not forward packets received with the broadcast address as the destination address at the link layer (which is the destination link-layer address of \ac{CBF} transmissions). This mechanism avoids multiple retransmissions and collisions at the border of the Destination Area that can happen when vehicles outside that GeoArea receive a packet from a vehicle that is actually inside the Destination Area, but \textcolor{black}{the vehicle} is not considered as such by its neighbors because its last position update (i.e., its last recorded \ac{CAM}) was when the vehicle was still outside the area. This situation creates Greedy retransmissions when many vehicles outside the Destination Area try to forward the packet to reach the GeoArea at the same time (although, in reality, the packet was already coming from there).

\item \textbf{Enabling source retransmissions}: This mechanism increases dissemination coverage. It works as follows: the vehicle that is the source of a packet proceeds to transmit it but also stores the packet in its own \ac{CBF} buffer. This mechanism improves the \ac{PDR} by reducing the packet loss that occurs when the transmission at the source coincides with transmissions from other nearby vehicles, generating a collision that prevents other vehicles from receiving the packet sent by the source. Note that, with the proposed mechanism in normal operation, the scheduled retransmission of the packet in the \ac{CBF} buffer of the source will \textcolor{black}{likely be} cancelled by the reception of a retransmission of the packet from another vehicle. 

\item \textbf{\acf{GPC}}: This mechanism increases the dissemination coverage by limiting the cancellation of the retransmission of packets in the \ac{CBF} buffer when receiving copies of the packets that are already in the buffer. With \ac{GPC}, a packet in the \ac{CBF} buffer is discarded only in cases where the sender of the copy of the packet is a better forwarder than the vehicle with the buffered packet. In the other case, \textcolor{black}{the timer associated with the packet in the \ac{CBF} buffer is updated according to the distance between the best previous sender and the ego vehicle}. This mechanism improves the \ac{PDR} by preventing a sender that does not represent any progress toward the border of the Destination Area from cancelling a retransmission from a sender that represents some progress (a simple example is two vehicles driving in parallel on a highway, the first vehicle retransmits a packet and, without \ac{GPC}, the second vehicle's transmission could cancel the retransmission on all vehicles that received the first vehicle's transmission). To check whether a new sender provides progress toward the border of the Destination Area, a simple formulation is used: to cancel the retransmission of a packet, a vehicle (Self) must check that the vehicle that sent the copy (Sender) is farther from the Source vehicle (origin of the packet) than Self, and that the distance between Sender and Source is greater than the distance between Sender and Self (this last condition helps to avoid cancellations when the Sender and Self vehicles are on opposite sides from the Source).

\item \textbf{\acf{FoT}}: This mechanism reduces unnecessary retransmissions caused by packets waiting in DCC queues. In \ac{FoT}, packets are kept in the \ac{CBF} buffer not only until the \ac{CBF} packet timer expires, but also until the \ac{DCC} gatekeeper is opened (i.e., the congestion control mechanism at the access layer allows sending a packet). This mechanism reduces retransmissions by keeping packets in the \ac{CBF} buffer while they cannot be transmitted, and reducing the time packets spend in the DCC queues (where they cannot be cancelled). Therefore, with \ac{FoT}, we have a longer time window in which it is possible to cancel the retransmission of a packet. The effects of \ac{FoT} are more noticeable with high network loads.  
\end{itemize}

In \cite{Amador2022} the performance of the proposed mechanisms was analyzed, although only in highway scenarios, and we concluded that, in these scenarios, the combination of the proposed mechanisms helped to achieve improved \ac{PDR}, using significantly fewer transmissions than ETSI ITS \ac{CBF}.

\subsection{Related work}

In the past some works \cite{Bellache:2017a, Kuhlmorgen2020, Turcanu2020, Spadaccino2020, riebl2021, Hajjej2022}, including ours (see above and \cite{Amador2022}), have discussed \ac{ETSI} GeoNetworking protocols, particularly \ac{CBF} as the area forwarding protocol, and proposed different solutions to overcome identified inefficiencies:
\begin{itemize}

\item \textbf{Reduce unnecessary retransmission caused by \ac{DCC} queues}. In the \ac{ETSI} architecture, once the \ac{CBF} timer expires, the \textcolor{black}{vehicle} becomes a forwarder, removes the packet from the \ac{CBF} buffer, and gives it to the access layer to be disseminated. However, the \ac{DCC} access layer may keep the packet for a while waiting in a \ac{DCC} queue before being transmitted. Our \ac{FoT} mechanism, presented above and in \cite{Amador2022}, is a proposal to address this very problem. The work in \cite{Kuhlmorgen2020} proposes RORA, a mechanism to eliminate duplicated packets from the \ac{DCC} buffer avoiding unnecessary retransmissions. However, RORA does not tackle the issue of selecting the best time-wise forwarder. As we will see later, this situation impairs \ac{CBF} performance. In addition, RORA is a cross-layer solution that requires a low layer (i.e., Access Layer \ac{DCC}) to realize the behavior and messages of an upper layer (i.e., \ac{CBF} at the Network Layer), making both layers tightly coupled.

\item \textbf{Reduce unnecessary retransmission caused by duplicated packets}. The \ac{ETSI} \ac{ITS} \ac{CBF} specification does not include a persistent duplicate packet detection mechanism. Several works in the literature, including ours (see above and \cite{Amador2022}), discuss the need for a persistent duplicate packet detection \ac{DPD} mechanism that keeps track of duplicates when the message is no longer at the \ac{CBF} buffer \cite{riebl2021}. A discussion on how to maintain the data structure for DPD can be found in \cite{riebl2021}.

\item \textbf{Increase dissemination coverage with more retransmissions.} A strategy to enhance the dissemination coverage consists of counting the number of copies received for a given message, and the candidate forwarders do not drop their CBF buffered messages until they are seen a certain number of times \cite{Torrent2007, Bellache:2017a, Spadaccino2020}. Additionally, in \cite{Bellache:2017a} is proposed to adapt this number of copies according to the channel load conditions. \textcolor{black}{Even if} \cite{Bellache:2017a} takes into account the existence of a \ac{DCC} mechanism at different layers, aligned with the \ac{ETSI} \ac{ITS} architecture, it does not consider \ac{DCC} at the Access Layer as \ac{ETSI} standardization does.

\item \textbf{Increase dissemination coverage: choosing the best forwarders.} In \cite{Hajjej2022}\textcolor{black}{, it }is proposed to modify the \ac{ETSI} \ac{CBF} protocol to combine \ac{CBF} (a receiver-based mechanism) with a sender-based approach where the current forwarder uses its neighbors' table and also includes in the packet a list of selected next forwarders, i.e., two \textcolor{black}{vehicles} spatially distributed around the current forwarder to cover a wider area. This approach tries to supplement the blind selection of forwarders that the \ac{ETSI} \ac{CBF} protocol does. This approach improves dissemination coverage and dissemination delay at the cost of increasing the number of transmissions in the network (i.e., overhead).

\item \textbf{Increase dissemination coverage: inhibitions from very close forwarders}. The works in \cite{Paulin2015} and our \ac{GPC} mechanism (see above and \cite{Amador2022}) tackle the problem of two close forwarders transmitting at almost the same time (without cancelling each other) which leads to the cancellation of all future forwarders, resulting in poor dissemination coverage. \cite{Paulin2015} proposes two possible solutions: (1) using a probabilistic cancellation of packets in the \ac{CBF} buffer, or (2) incorporating a progress check, with the idea that a packet cancels a packet in the \ac{CBF} buffer only if the new forwarder represents more progress to the destination than itself, otherwise the \ac{CBF} timer is updated according to the position of the new transmitter. However, this work only discusses CBF as a non-area forwarding mechanism (i.e., to reach a distant Destination Area), and it does not consider the use of a \ac{DCC} mechanism or delivery to the application layer. 

\item \textbf{Adapt \ac{CBF} to specific vehicular applications: \ac{FCD} service } The work in \cite{Turcanu2020} proposes rCBF a modified version of \ac{ETSI} \ac{ITS} \ac{CBF} tailored for \ac{FCD} service. rCBF pursues two main objectives, namely, avoiding simultaneous retransmissions and reducing the number of next-forwarders retransmissions. To this end, the calculation of the \ac{CBF} Timer ($T_{CBF}$) has been modified: a low-weight random term has been incorporated into the \ac{CBF} Timer calculation, and vehicles further away from the sender than $DIST_{MAX}$ refrain from retransmitting packets (i.e. the \ac{CBF} Timer is set to infinite).

\end{itemize}

A relevant issue to highlight is that just a few works proposing enhancements to \ac{ETSI} \ac{ITS} \ac{CBF} protocol have actually considered the complete \ac{ETSI} \ac{ITS} architecture in the experimental evaluation: \cite{riebl2021, Kuhlmorgen2020} and ours (as described above and in \cite{Amador2022}). In particular, it is common that these works~\cite{Bellache:2017a, Turcanu2020, Hajjej2022} do not consider the \ac{DCC} at the Access Layer, which may have a non-negligible impact on dense networks.

\ac{ETSI} \ac{ITS} \ac{CBF} aims to be an area-forwarding mechanism to be used both on highways and in urban scenarios. An open issue is the evaluation of the \ac{ETSI} \ac{ITS} \ac{CBF} protocol and its enhancements in urban scenarios, since many of these works, such as \cite{Kuhlmorgen2020} and our previous work~\cite{Amador2022}, only consider highway scenarios, or mixed scenarios that combine different road categories (e.g., residential, arterial, and highway)~\cite{Hajjej2022, Turcanu2020}.

\section{Further improvements to ETSI ITS CBF}
\label{sec:improvements}

\subsection{Slotted CBF}
\label{subsec:scbf}

\begin{figure}[h]
	\centering
	\includegraphics[width=\textwidth]{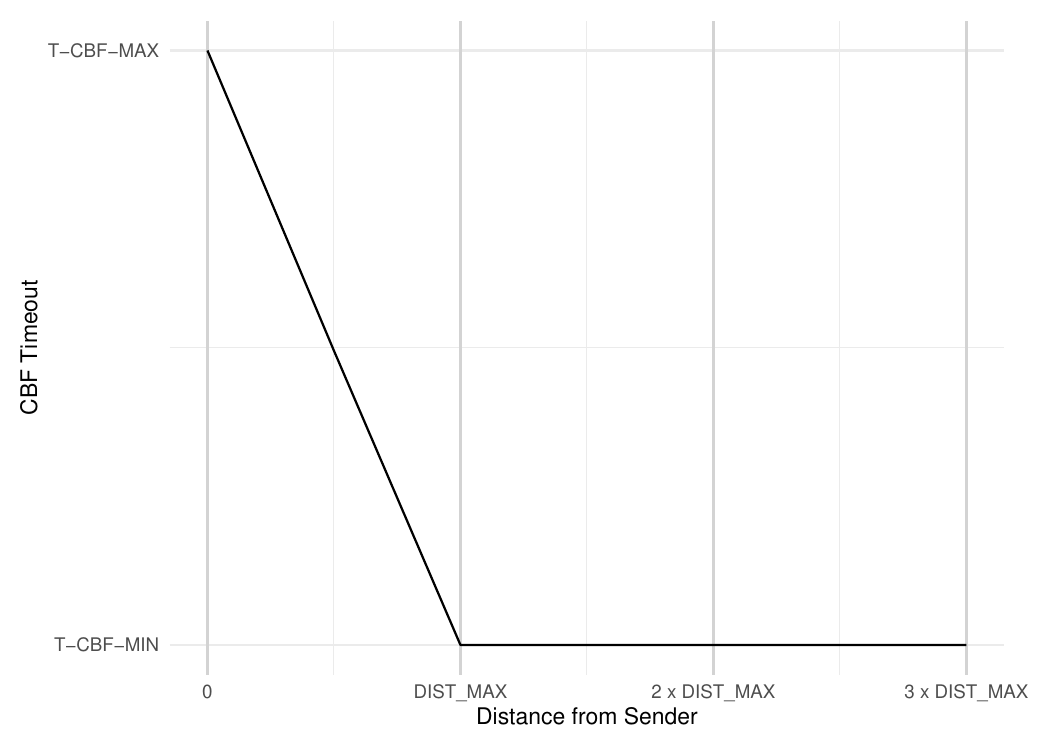}
	\caption{ETSI CBF Timeout calculation}
	\label{fig:cbf}
\end{figure}

The \ac{ETSI} \ac{CBF} algorithm calculates the \ac{CBF} timer, $T_{CBF}$, for a received packet when it is inserted in the \ac{CBF} buffer using equation~\ref{eq:TempCBF}:

\begin{equation}
\label{eq:TempCBF}
T_{CBF} = \begin{cases}
T_{CBF-MAX} - \frac{T_{CBF-MAX} - T_{CBF-MIN}} {DIST_{MAX}} \times DIST &   \text {if $DIST$ $\leq$ $DIST_{MAX}$} ,\\
T_{CBF-MIN} &   \text {if $DIST$ $>$ $DIST_{MAX}$}
\end{cases}
\end{equation}
where $T_{CBF-MAX}$ is the maximum value of the \ac{CBF} timer (default value 100~ms), $T_{CBF-MIN}$ is the minimum value of the \ac{CBF} timer (default value 1~ms), $DIST_{MAX}$ is the maximum communication range of the used radio technology (default value 1000~m), and $DIST$ is the distance between the vehicle that has received the packet and the previous forwarder). Figure~\ref{fig:cbf} represents the values of the \ac{CBF} timer for different values of $DIST$ resulting from applying equation~\ref{eq:TempCBF}. In this figure, the effect of assuming a certain maximum communication range, $DIST_{MAX}$, in the used wireless access technology is shown: all vehicles at distances greater than $DIST_{MAX}$ end up using the same \ac{CBF} timer for the retransmission of the packet. 

In our experiments, we have detected situations where receptions occurred beyond 1000~m (the default value of $DIST_{MAX}$ in \ac{ETSI} \ac{CBF}). Although the \ac{PDR} is low at those distances, some vehicles can still receive and retransmit packets at them. Therefore, collisions occur when several vehicles try to retransmit a packet at exactly the same time, and this could affect the best potential forwarders at the edge of the communications range.

A potential simple solution may be to increase the value of $DIST_{MAX}$ and the related maximum \ac{CBF} timer ($T_{CBF-MAX}$). However, this requires being able to determine the appropriate $DIST_{MAX}$ for any situation, even considering future scenarios where \ac{CBF} can be applied with mixed ITS-G5 and C-V2X~\cite{Soto2022} radio technologies. Note that it is not enough to choose an arbitrarily large $DIST_{MAX}$, because the $T_{CBF-MAX}$ has to be adapted as well, and this would increase packet forwarding delays. 

Another possible solution, proposed in \cite{Turcanu2020}, is to configure vehicles not to retransmit packets received further away than $DIST_{MAX}$ (i.e., the \ac{CBF} timer in equation~\ref{eq:TempCBF} is $\infty$ when $DIST > DIST_{MAX}$). However, there may be situations where this may result in lost packets that could have been distributed further away in the area of interest (e.g., when there are not any receivers at shorter distances).

Therefore, we propose a new mechanism to use with \ac{CBF} that we call Slotted CBF (\mbox{S-CBF}). In \mbox{S-CBF}, to calculate the \ac{CBF} timer, vehicles first calculate their slot from the transmitter using equation~\ref{eq:S-CBF}. 

\begin{equation}
\label{eq:S-CBF}
Slot_{number}=\lceil \frac{DIST}{DIST_{MAX}} \rceil
\end{equation}

Then, the vehicle calculates the \ac{CBF} timer using equation~\ref{eq:Temp_S-CBF}. 

\begin{equation}
\label{eq:Temp_S-CBF}
T_{CBF} = 
T_{CBF-MAX} \times Slot_{number} - \frac{T_{CBF-MAX} - T_{CBF-MIN}} {DIST_{MAX}} \times (DIST - DIST_{MAX} \times (Slot_{number} - 1)) 
\end{equation}

\begin{figure}[t]
	\centering
	\includegraphics[width=\textwidth]{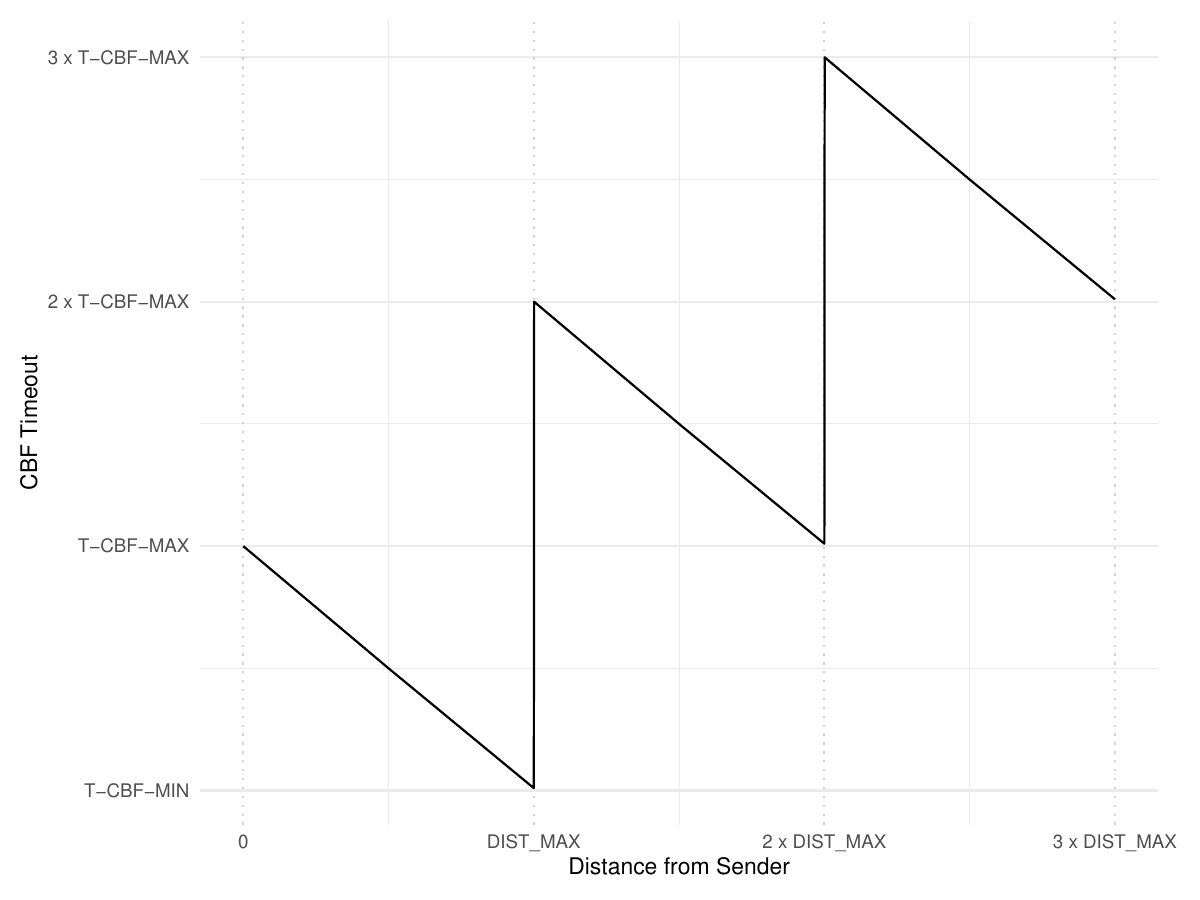}
	\caption{Slotted-CBF (S-CBF) Timeout calculation}
	\label{fig:s-cbf}
\end{figure}

The rationale is that at distances from the last forwarder less than $DIST_{MAX}$ (Slot 1), equation~\ref{eq:Temp_S-CBF} is equal to equation~\ref{eq:TempCBF}, and S-CBF performs in the same way as ETSI \ac{CBF}. Therefore, the vehicles with shorter \ac{CBF} timers are those at distances close but less than $DIST_{MAX}$, and thus they are the ones favored to retransmit the packet. Notice that preferring a forwarder beyond $DIST_{MAX}$ may lead to vehicles near the previous sender not cancelling their copies and thus increasing the number of transmissions. At distances greater than $DIST_{MAX}$ and less than $2 \times DIST_{MAX}$ (Slot 2) the philosophy is repeated, but with timers between $T_{CBF-MAX}$ and $2 \times T_{CBF-MAX}$, so vehicles at Slot 1, if available, will transmit before vehicles in Slot 2, but vehicles in Slot 2 still can transmit when no vehicle transmits in Slot 1. The same happens for further slots. The values of the \ac{CBF} timer at different distances applying \mbox{S-CBF} are shown in Fig.~\ref{fig:s-cbf}. 

 When a vehicle receives a packet that is already in its \ac{CBF} buffer, by applying \ac{GPC}~\cite{Amador2022}, the packet in the \ac{CBF} buffer must not be discarded when the retransmitter of the packet is not a better forwarder than the vehicle with the buffered packet. In that case, only the \ac{CBF} timer must be updated to reflect the new transmitter. In \mbox{S-CBF}, \textcolor{black}{the timer is updated using equation~\ref{eq:Update_temp_S-CBF}:}

\begin{equation}
\label{eq:Update_temp_S-CBF}
T_{CBF_{new}} =  \begin{cases}
max \{ {T_{CBF}}_{buffer}, {T_{CBF}}_{received\_packet} \} & \text {if within the same slot,}\\  
min \{ {T_{CBF}}_{buffer}, {T_{CBF}}_{received\_packet} \} & \text {if the slot has changed}
\end{cases}
\end{equation}
where ${T_{CBF}}_{buffer}$ is the \ac{CBF} timer associated with the packet in the \ac{CBF} buffer, and ${T_{CBF}}_{received\_packet}$ is the \ac{CBF} timer for the newly received copy of the packet calculated using equation~\ref{eq:Temp_S-CBF}. \textcolor{black}{In equation~\ref{eq:Update_temp_S-CBF}, there are two cases depending on whether the vehicle receiving the retransmission is in the same slot from the new transmitter as it was from the transmitter of the copy of the packet in the CBF buffer.} When the slot is the same, the new \ac{CBF} timer must be the maximum of the timer of the packet in the \ac{CBF} buffer and the timer calculated for the newly received copy of the packet, since the value of the new timer must correspond to the better of the two forwarders (the better forwarder is the closest, which means a longer timer). With this value of the \ac{CBF} timer, the vehicle is yielding the opportunity to retransmit the packet to possible better forwarders that received the original packet or the retransmission. When the slot corresponding to the transmission of the packet in the buffer is different from the slot corresponding to the newly received copy of the packet, the new timer must be the minimum of the two timers. The reason is to adapt the timer to the time band corresponding to the best slot of the two transmissions.

\subsection{\acf{FoT+}}

This mechanism is an improvement of the \ac{FoT} algorithm we proposed in \cite{Amador2022}. \ac{FoT} tries to maximize the time that a packet stays in the \ac{CBF} buffer, and thus the probability of being cancelled by a retransmission. The problem with the original \ac{ETSI} \ac{CBF} mechanism is that, when a packet is extracted from the \ac{CBF} buffer, it is passed to a \ac{DCC} queue (there are four \ac{DCC} queues for packets with different priorities: TC0, TC1, TC2, TC3), where it should wait until the \ac{DCC} gate opens the gate and sends the highest-priority packet in the \ac{DCC} queue (TC0 has the highest priority and TC3 has the least one). This waiting time at the \ac{DCC} queue may be significant if the vehicle has just sent another packet (e.g., a \ac{CAM}), especially in congested areas, where the time between consecutive packets of the same vehicle may be up to 1000~ms. However, since the forwarded packet is waiting in a \ac{DCC} queue instead of the \ac{CBF} buffer, it cannot be cancelled even if another copy of the packet has been received.

RORA~\cite{Kuhlmorgen2020} proposed solving this problem with a cross-layer mechanism where the DCC gate is aware of \ac{CBF} packets (i.e., \ac{DCC} in the access layer must be able to interpret GeoNetworking headers), so it can drop the repeated \ac{CBF} packets (by checking the \ac{CBF} buffer \ac{DPL}) when they are dequeued from a \ac{DCC} queue.

\ac{FoT}~\cite{Amador2022} is a cleaner mechanism from an architectural point of view (notice that all GeoNetworking protocols use the \ac{DCC} access layer to send their packets): it only requires \ac{CBF} to be aware of the time when the \ac{DCC} gate is going to be open (\ac{DCC} provides feedback to upper layers of the ETSI architecture through the Management Plane). Thus, in \ac{FoT} the timer associated with a packet in the \ac{CBF} buffer is the maximum between the regular distance-based ${T_{CBF}}_{received\_packet}$ and the time until the \ac{DCC} gate will open again ($t_{DCC}$), as illustrated in \textcolor{black}{equation~\ref{eq:fot}}.

\begin{equation}
\label{eq:fot}
{T_{CBF}}_{FoT} =  max \{T_{CBF}, t_{DCC}\}
\end{equation}

In case $t_{DCC}$ changes before the \ac{CBF} timeout expires (e.g. if another packet has been sent meanwhile), if the DCC gate is not open when the timer expires ($t_{DCC} > 0$), it is rescheduled again for $t_{DCC}$, so it can wait for the DCC gate to be open at the \ac{CBF} buffer instead of a \ac{DCC} queue. In congested highway scenarios, \ac{FoT} allows a higher number of \ac{CBF} packets to be cancelled, and thus reduces the total number of transmissions \cite{Amador2022}.

A limitation of the original \ac{FoT} mechanism is that it does not always guarantee that the forwarded \ac{CBF} packet is transmitted directly, without waiting at the \ac{DCC} queue (TC3). This may happen because a high-priority packet (e.g., a \ac{CAM} or a local \ac{DENM}) is already waiting at the \ac{DCC} queues (e.g. TC2 or TC0/TC1, respectively) when the \ac{DCC} gate opens. Thus, the \ac{DCC} scheduler chooses the high-priority packet instead of the forwarded \ac{CBF} packet, which has to wait in the \ac{DCC} queue until the \ac{DCC} gate opens again, thus it cannot be cancelled by another transmission because it is no longer at the \ac{CBF} buffer.

Luckily, the solution is fairly simple, instead of waiting exactly until the \ac{DCC} gate opens, with \ac{FoT+} the \ac{CBF} packet waits in the \ac{CBF} buffer a little more (e.g. $\epsilon = 1~ms$) than with \ac{FoT} and then checks the DCC gate\textcolor{black}{. Therefore, the timer associated with the packet in the CBF buffer is determined by equation~\ref{eq:fot+}}. 
\begin{equation}
\label{eq:fot+}
{T_{CBF}}_{FoT+} =  max \{T_{CBF}, t_{DCC} + \epsilon\}
\end{equation}

\textcolor{black}{For the case when, after $\epsilon$, the gate} is still open ($t_{DCC}=0$), it means that no other packet was waiting at the \ac{DCC} queues and thus the \ac{CBF} packet can be transmitted directly. On the other hand, if the \ac{DCC} gate is closed, this means that another packet has been just sent, and the CBF packet must keep waiting at the CBF buffer for $t_{DCC} + \epsilon$ until the \ac{DCC} gate opens again, repeating the whole process. \textcolor{black}{FoT+ does not change the time when packets are sent, except for the small $\epsilon$, but only where they wait. Thanks to FoT+, when the CBF timer of a packet expires, if a higher priority packet (e.g., a CAM) is waiting to be sent in the corresponding DCC queue, the packet in the CBF buffer is kept there, where it can be cancelled, instead of being sent to wait in a DCC queue. In FoT, the packet in the CBF buffer is kept there until the DCC gate is opened, but then the packet is sent to the DCC queue and will have to wait there because the higher priority packet has to be sent and consumes the DCC turn.}   

It is worth noting that this solution is only feasible when no other TC3 traffic exists in the \textcolor{black}{vehicle} (i.e., other than \ac{DENM} and \ac{CAM} traffic). Otherwise, the forwarded CBF packet will always cede its turn in the TC3 queue to the local TC3 traffic, which may lead \textcolor{black}{to longer delays and even starvation of forwarded traffic}.

\begin{figure}[tbh!]
	\centering
	\includegraphics[width=\textwidth]{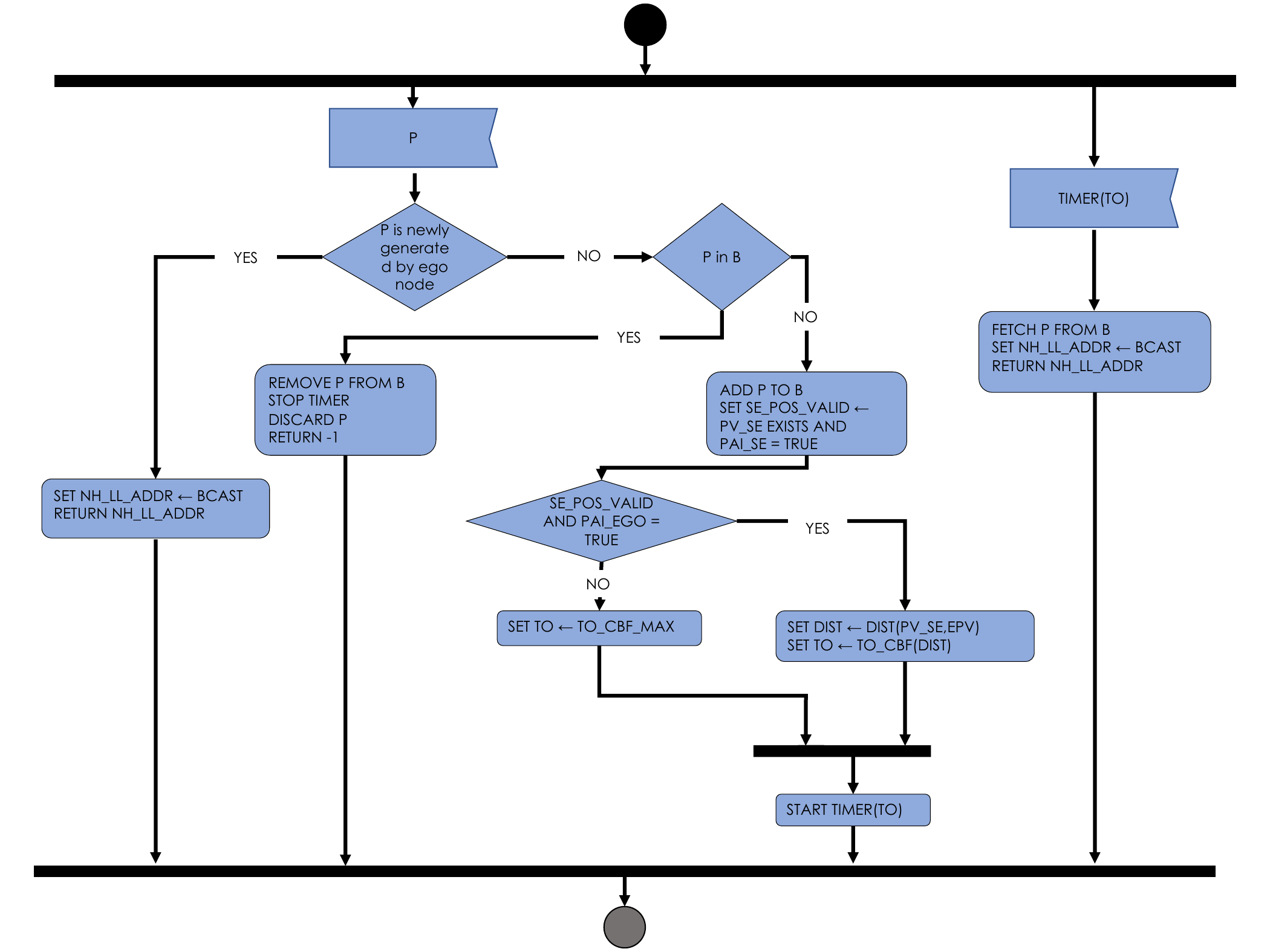}
        \caption{Activity diagram for ETSI CBF (Adapted from Annex F.3 in~\cite{etsiNewGeoNetworking})}
        \label{fig:diagram_etsi_cbf}
\end{figure}

\begin{figure}[tbh!]
	\centering
	\includegraphics[width=\textwidth]{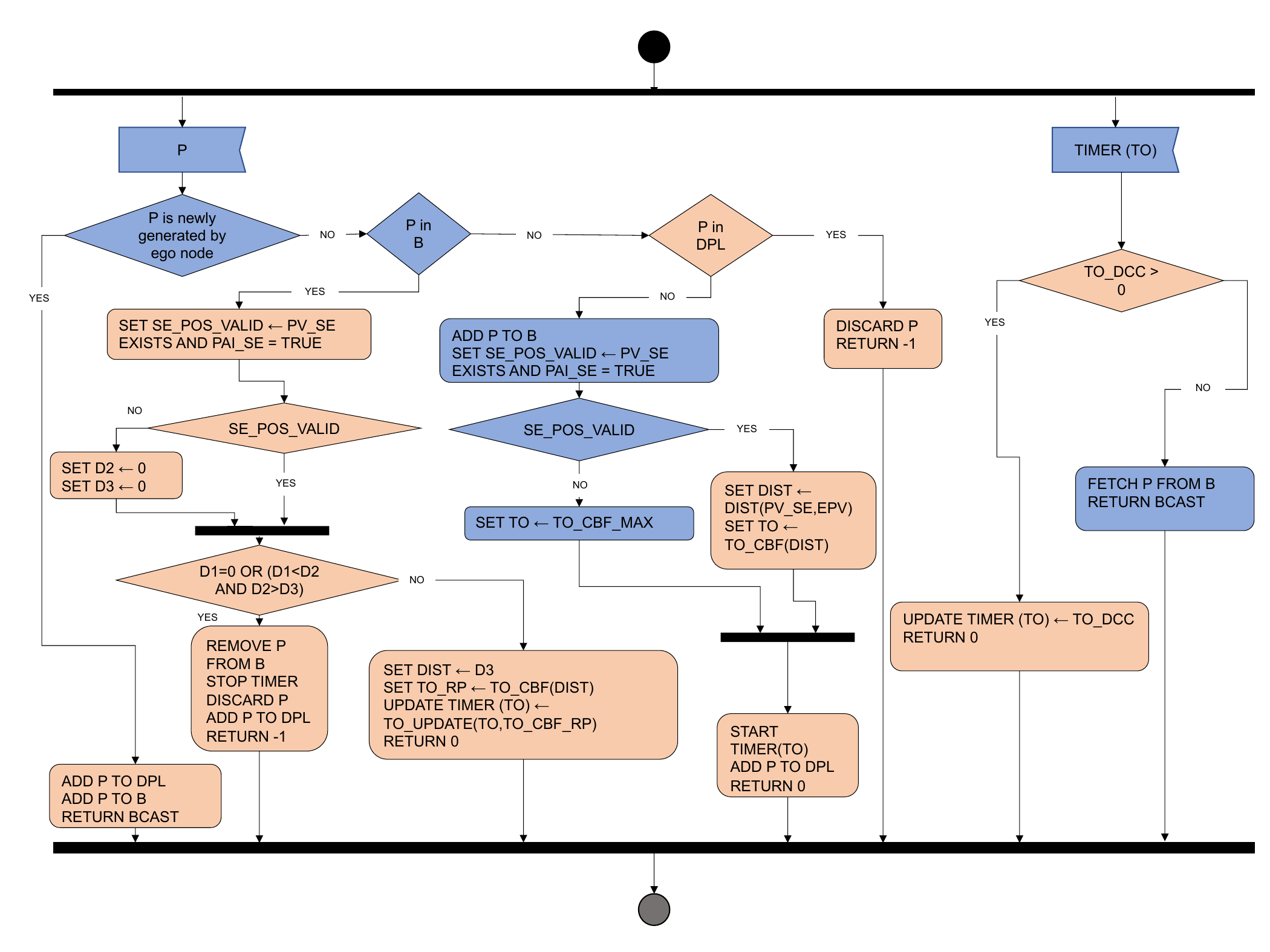}
        \caption{Activity diagram for FoT variants}
        \label{fig:diagram_fot}
\end{figure}

\textcolor{black}{Figs.~\ref{fig:diagram_etsi_cbf} and~\ref{fig:diagram_fot} show the activity diagrams for ETSI \ac{CBF} and the FoT variants. The differences are highlighted on Fig.~\ref{fig:diagram_fot}. The process for calculating and updating timers, TO\_CBF(DIST) in the diagram, uses the equations described in Section\ref{sec:improvements} depending on which variant of FoT is used (i.e., FoT, S-FoT, or S-FoT+), e.g., Equation~\ref{eq:S-CBF} is only used for S-FoT and S-FoT+. Furthermore, Fig.~\ref{fig:diagram_fot} also highlights other improvements included in FoT variants, such as duplicate-packet detection, enabling source retransmission, and geographically-aware packet cancellations.}

\section{Evaluation in highway scenarios}
\label{sec:highway}

\begin{table}[h]
	\centering
	\caption{Simulation Parameters}
	\label{tbl:simpars}
	\begin{tabular}{| l | l |}
		\hline
		\textbf{Parameter}  & \textbf{Values} \\
		\hline
		Access Layer protocol & ITS-G5 (IEEE 802.11p) \\
		Channel bandwidth & 10\,MHz at 5.9\,GHz \\
		Data rate & 6\,Mbit/s \\
		Transmit power & 20\,mW \\
		Path loss model & Two-Ray interference model\textcolor{black}{~\cite{Sommer:2012}} \\
		Maximum transmission range measured & 1500\,m \\
		CAM packet size & 285 bytes \\
            \textcolor{black}{CAM generation frequency} & \textcolor{black}{Variable rate 1--10~Hz (ETSI CAM~\cite{etsiCA})} \\
		CAM Traffic Class & TC2 \\
		DENM packet size & 301 bytes \\
		DENM Traffic Class & TC0 (Source) and TC3 (Forwarders) \\
		DENM lifetime & 10\,s\\
		DPL size & 32 packet identifiers per Source\\
		\hline
		\multicolumn{2}{| c |}{\textit{Urban Scenario}}\\\hline
		Obstacle model & Simple obstacle shadowing\\\hline
		Obstacle parameters & Buildings: 9\,dB/cut, 0.4\,dB/m\\\hline
	\end{tabular}
\end{table}

We evaluate the performance of \ac{ETSI} Contention-Based Forwarding \textcolor{black}{(ETSI CBF)}, \ac{ETSI} Simple GeoBroadcast forwarding \textcolor{black}{(for simplicity, labeled ETSI Simple in the figures and tables below)}, our previous proposals~\cite{Amador2022}, and the two new improvements proposed in this paper. We perform a set of experiments using the Artery~\cite{Artery} simulation toolkit. Artery, which runs on OMNET++, implements the \ac{ETSI} ITS-G5 protocol stack through Vanetza, an open-source C++ implementation of the ETSI specification which includes GeoNetworking and \ac{DCC}. \textcolor{black}{Artery uses Veins~\cite{Veins} to implement 802.11p for the MAC sublayer and the physical layer}. For our experiments, we use the Artery implementations of GeoNetworking and \ac{DCC} (i.e., \ac{CBF} and Simple GeoBroadcast forwarding, and the \ac{ETSI} Adaptive \ac{DCC} approach~\cite{etsiNewDcc}), and we implement all our proposals (i.e., \ac{FoT} variants) on top of them. Notice that all improvements are cumulative: \ac{FoT} refers to all mechanisms defined in \cite{Amador2022}, S-FoT refers to the Slotted CBF mechanism built on top of \ac{FoT}, and S-FoT+ further adds FoT+ to S-FoT. Artery uses SUMO (Simulation of Urban MObility)~\cite{sumo2012} to simulate vehicle mobility. In our two deployments, highway and urban, we use the default configuration for vehicle behavior and we only include passenger vehicles. Parameters for the simulations are shown in Table~\ref{tbl:simpars}, and details pertaining to each scenario are described in their particular sections. All simulations have been repeated 5 times with different random seeds, and the result values are the mean of the 5 runs. \textcolor{black}{Plots show mean values with error bars representing 95\% confidence intervals.}

\subsection{Simulation scenario}
\label{subsec:sim_highway}
For the highway scenario, we simulate a 5\,km road with four lanes in each direction. On the west end of the road, we locate a stationary vehicle in the median that represents a broken-down car. The stationary vehicle starts sending \acp{DENM} to a rectangular shape that covers both road directions up to 4\,km to the east. These \acp{DENM} are generated at a 1\,Hz frequency and they are meant to be forwarded (i.e., sent as GeoBroadcast as opposed to Single-hop Broadcast). \textcolor{black}{Other vehicles drive along the road in both directions, and we have experiments with different densities: 10, 20, 30, 40, and 50 vehicles/km per lane, in order to measure the effect of network congestion. The movement of the vehicles is controlled by SUMO~\cite{sumo2012}. All vehicles send CAM traffic according to ETSI rules~\cite{etsiCA}.}

\subsection{Results}
\label{subsec:results_highway}

Our evaluation is based on the following three performance metrics:
\begin{itemize}
    \item \textbf{Number of transmissions}: number of times DENMs are transmitted or retransmitted in the scenario. 30 DENMs are transmitted by the source vehicle (one time per second for 30 seconds), and the rest of the transmissions correspond to retransmissions in other vehicles to cover the Destination Area. 
    \item \textbf{\acf{PDR}}: the ratio of vehicles that received a given DENM to the number of vehicles within the Destination Area. Our PDR measurements also account for vehicles that depart or enter the area while the DENM is being forwarded, and thus PDR may rise above 100\% (e.g., new vehicles enter the area while the message is being forwarded so more vehicles receive the message than vehicles were present in the area when the message was sent).
    \item \textbf{End-to-end delay}: time elapsed since a DENM is \textcolor{black}{generated at the facilities layer in} the source vehicle until it is successfully received \textcolor{black}{for the first time at the facilities layer of a} vehicle within the Destination Area. \textcolor{black}{For each transmitted packet, we collect a sample of end-to-end delay for each receiving vehicle.}  
\end{itemize}

\begin{figure}[tbh!]
    \centering
    \includegraphics[width=0.8\textwidth]{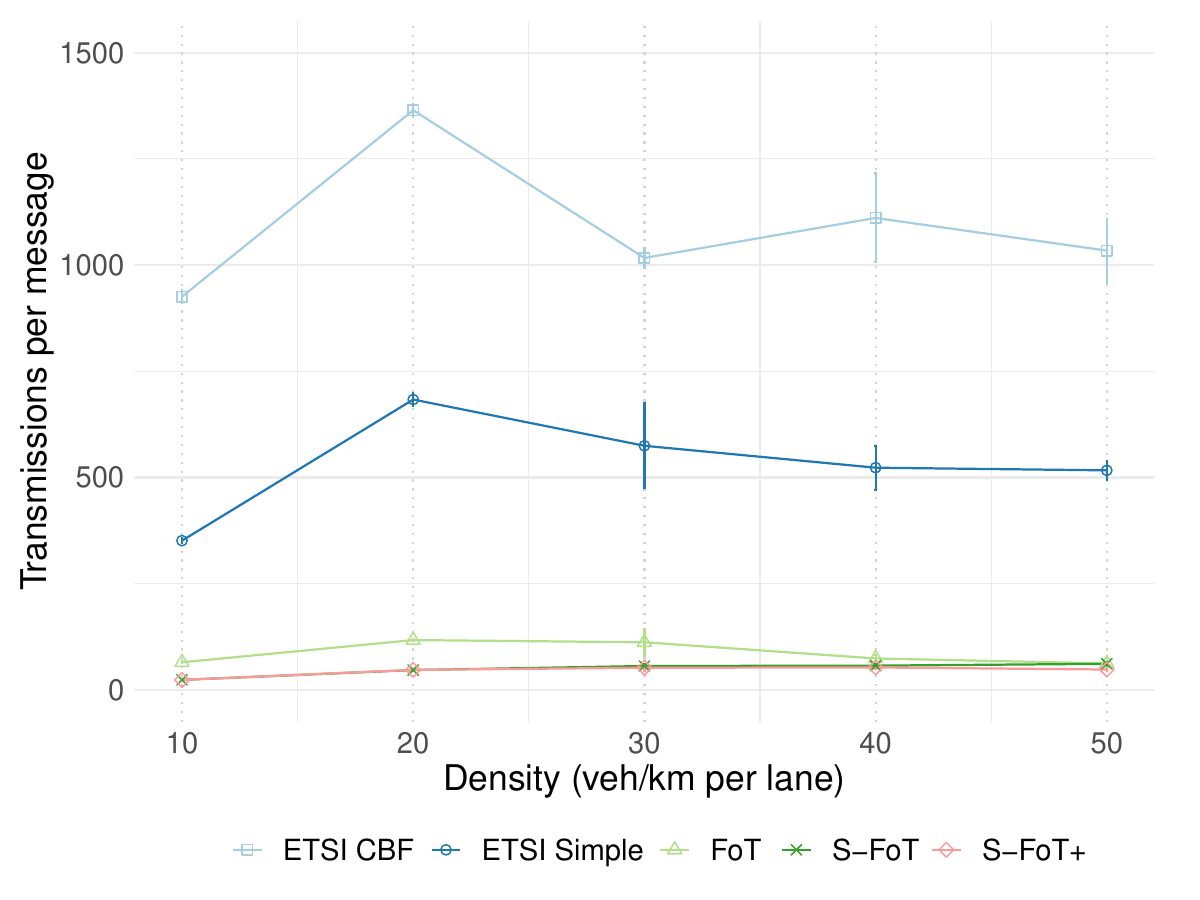}
    \caption{Transmissions per density in the highway scenario}
    \label{fig:tx_highway}
\end{figure}

Fig.~\ref{fig:tx_highway} shows the difference between the number of transmissions multi-hop mechanisms produce. Table~\ref{table:hwy_pdr} shows that the difference in transmissions does not reflect in a higher \ac{PDR}, concurring with~\cite{Amador2022}, where network congestion awareness helped FoT to reduce transmissions in an order of magnitude. Furthermore, the table reflects another important result: slotted mechanisms (i.e., S-FoT and S-FoT+) have a significant improvement in transmission efficiency for reasons that we explain in the upcoming paragraphs.

\begin{table*}[tbh!]
\caption{Average Packet-Delivery Ratio (PDR) for the highway scenario}
\centering
\begin{tabularx}{\textwidth}{| >{\raggedright\arraybackslash}X |  >{\raggedleft\arraybackslash}X |
>{\raggedleft\arraybackslash}X | >{\raggedleft\arraybackslash}X | >{\raggedleft\arraybackslash}X | >{\raggedleft\arraybackslash}X |}
\hline
\textbf{Density (veh/km per lane)} & \textbf{ETSI CBF} & \textbf{ETSI Simple} & \textbf{FoT} & \textbf{S-FoT} &\textbf{S-FoT+} \\\hline
10 & 0.9999  & 0.9934  & 0.9999  & 0.9999  & 1.0000 \\ \hline
20 & 1.0023  & 1.0005  & 1.0006  & 1.0006  & 1.0006 \\ \hline
30 & 1.0023  & 1.0013  & 1.0011  & 1.0017  & 1.0011 \\ \hline
40 & 1.0002  & 1.0012  & 1.0026  & 1.0031  & 1.0046 \\ \hline
50 & 0.9954  & 0.9839  & 1.0059  & 0.9996  & 1.0056 \\ \hline
\end{tabularx}
\label{table:hwy_pdr}
\end{table*}

\begin{table*}[tbh!]
\caption{Average number of transmissions \textcolor{black}{per message} for the highway scenario}
\centering
\begin{tabularx}{\textwidth}{| >{\raggedright\arraybackslash}X |  >{\raggedleft\arraybackslash}X|  >{\raggedleft\arraybackslash}X | >{\raggedleft\arraybackslash}X | >{\raggedleft\arraybackslash}X | >{\raggedleft\arraybackslash}X | >{\raggedleft\arraybackslash}X |}
\hline
\textbf{Density (veh/km per lane)} & \textbf{Avg. veh. in dest. area} &\textbf{ETSI CBF} & \textbf{ETSI Simple} & \textbf{FoT} & \textbf{S-FoT} &\textbf{S-FoT+} \\\hline
10 & 338.80 & 925.60  & 351.38 & 65.11 &   23.64 &   23.66\\ \hline
20 & 632.20 & 1,364.42 & 683.50 & 117.38 & 46.85 & 47.24\\ \hline
30 & 957.40 & 1,016.72 & 574.66 & 112.22 & 56.33 & 52.25\\ \hline
40 & 1,151.80 & 1,110.89 & 523.06 & 73.13 & 57.21 & 52.90\\ \hline
50 & 1,378.20 & 1,033.97 & 516.84 & 62.43 & 62.43 & 48.09\\ \hline
\end{tabularx}
\label{table:highway_txr}
\end{table*}

\begin{figure}[tbh!]
    \centering
    \includegraphics[width=0.8\textwidth]{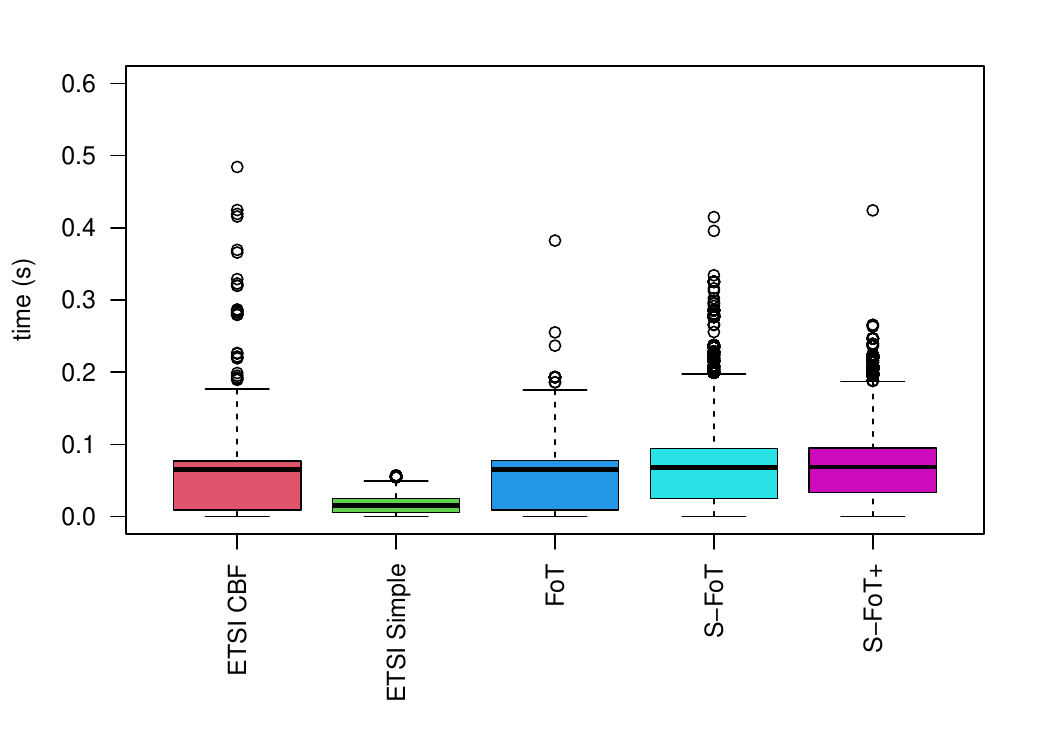}
    \caption{End-to-end delay for the 10 veh/km per lane density}
    \label{fig:e2e}
\end{figure}

The effect of synchronization caused by forwarders beyond $DIST_{MAX}$ is visible in Table~\ref{table:highway_txr} and Fig.~\ref{fig:tx_highway}. At low and medium densities, slotted mechanisms transmit less than half when compared to \ac{FoT}. An analysis of traces shows that forwarders beyond the maximum distance try to transmit at the same time, causing collisions and effectively negating the advantage of receiving at a long distance. Fig.~\ref{fig:e2e} shows that the slotted mechanisms do not add significantly to end-to-end message latency, with the median values of S-FoT and S-FoT+ at the same level of \ac{FoT} and even standard \ac{ETSI} \ac{CBF}. Latency for the \ac{ETSI} Simple GeoBroadcast mechanism is lower due to the nature of the algorithm, where all  \textcolor{black}{vehicles} try to forward packets immediately upon receipt.

These results are explained by the fact that the slotted mechanisms always take advantage of receptions by forwarders beyond $DIST_{MAX}$. In non-slotted mechanisms, these forwarders are only effective if there are not any other neighbors present, causing collisions that, in practice, make forwarders within $DIST_{MAX}$ effectively relay the message for most of the cases. Thus, messages cover the area in relatively the same amount of time, but slotted mechanisms use the medium more efficiently (i.e., lowering the number of ineffective messages).

Finally, as exhibited in Table~\ref{table:highway_txr} and Fig.~\ref{fig:tx_highway}, slotted mechanisms \textcolor{black}{maintain a constant performance across all densities (having more vehicles does not result in a significant increase in transmissions to cover the Destination Area)}. Starting at the 20~veh/km per lane density, the average number of transmissions for \mbox{S-FoT+} stays in a range within \textcolor{black}{6 transmissions of difference between the minimum and maximum value (47.24 -- 52.90)} for the remaining densities. This is an indicator that the mechanisms included in the algorithm reduce significantly the number of transmissions, enable an effective cancellation of buffered packets, and counteract the phenomena that destabilize the contention mechanism (i.e., \ac{DCC} and cancellation by sub-optimal forwarders).

Results for S-FoT+ show that \ac{DCC} and its queuing/de-queuing process have a significant effect on the efficiency of the contention mechanism. While FoT by itself tries to counteract unnecessary queuing and maximize the chances for other contenders to forward a message efficiently (i.e., using geographically-aware cancellation), the coexistence of different types of traffic within a  \textcolor{black}{vehicle} causes a phenomenon where, if a higher-priority message is also waiting at the \ac{DCC} queue, the contending packet will yield, and it will either be transmitted untimely or expire at the queue. Table~\ref{table:highway_txr} shows that this effect is more evident with higher channel occupation, since the difference between S-FoT+ and S-FoT is bigger than the difference between FoT and S-FoT, thus, indicating that it is ineffective queuing rather than receptions beyond $DIST_{MAX}$ that affect efficiency in congested scenarios, where receptions at long distances are less likely to occur.

\begin{figure}[tbh!]
    \centering
    \includegraphics[width=0.8\textwidth]{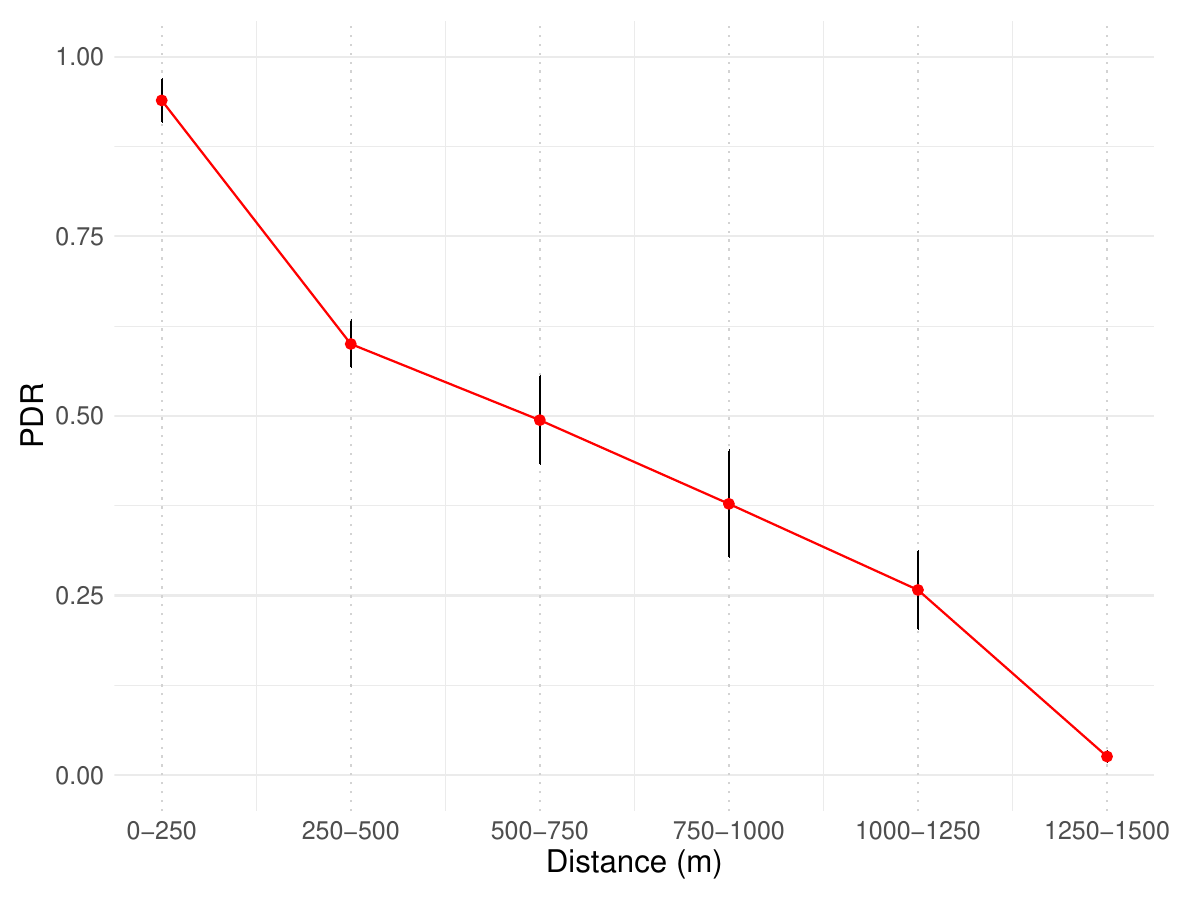}
    \caption{Packet-Delivery Ratio for ETSI SHB in a density of 30 veh/km per lane}
    \label{fig:pdr_shb_highway}
\end{figure}

As an additional note, Fig.\ref{fig:pdr_shb_highway} shows the \ac{PDR} for ETSI \ac{SHB} in a density of 30 veh/km per lane. In this scenario, the 30 \acp{DENM} are only sent from the source \textcolor{black}{vehicle} and are not intended to cover a particular Destination Area but just to reach as many neighbors as possible without the help of forwarders. The first segment, from 0 to 250\,m, has an average success rate of 94\%, which decreases to 60\% for the range between 250 and 500\,m. Losses due to attenuation bring success rates down along the distance. However, reachability can be increased by using multi-hop broadcast mechanisms, as we have seen in this section, and the cost in the number of transmissions is relatively low using S-FOT+. Another point to highlight is that there are still some successful receptions at distances above 1000\,m, which reinforce the need to consider forwarders beyond $DIST_{MAX}$ as potential causes for collisions in non-slotted mechanisms.

The outcomes from the highway scenario provide three main takeaways: 1) mechanisms that consider the effect of \ac{DCC} and geographical-induced cancellations outperform current standardized approaches, 2) the synchronization of forwarders at distances longer than $DIST_{MAX}$ adds up to ineffective forwarding, and 3) S-FoT+ shows more stability along a wide range of densities.

\section{Evaluation in urban scenarios}
\label{sec:urban}

One of the main contributions of this work is the evaluation of the ETSI GeoNetworking standards in urban scenarios. In a city, the effect of obstacles and road topology does not only affect signal propagation, but also the goal of broadcasting a message. For instance, in a city, two vehicles --A and B-- might be within a short distance from each other but with a building in between, while another vehicle --C-- might be at a farther distance but on the same street as A, which complicates the definition of "neighborhood" between vehicles A, B, and C in urban settings. In this section, we first evaluate the performance of \ac{ETSI} \ac{CBF} and \ac{ETSI} Simple GeoBroadcast forwarding, and compare it to \ac{FoT}-based mechanisms in an urban setting. Then, we evaluate the performance of multi-hop broadcast mechanisms (i.e., \ac{ETSI} \ac{CBF}, \ac{ETSI} Simple GeoBroadcast, FoT, and S-FoT+) against ETSI \ac{SHB} in order to assess the appropriateness of each mechanism to disseminate emergency messages in urban environments.

\subsection{Simulation scenario}
\label{subsec:sim_urbany}
To evaluate the forwarding mechanisms, we have employed the same dataset as in \cite{Uruena2017}. We use a map of central Madrid obtained from Open Street Map \cite{OSM} that contains long, multi-lane avenues, as well as medium-sized and small streets. The maximum speed of all streets has been set to 50 km/h. The map is a square of 4~km per side, allowing for a Destination Area shaped as a circle of 10~km$^2$, the largest GeoArea allowed by the ETSI standard~\cite{etsiNewGeoNetworking}.

\textcolor{black}{The movement of vehicles is controlled by SUMO~\cite{sumo2012}. Each vehicle makes a trip with a start and end point. Trips are randomly generated and initiated, and are specified so they have a minimum trip length of 1~km. The routes to perform the trips are calculated using the Dynamic User Assignment (DUA) \cite{Gawron1999} and A-star \cite{a-star} algorithms, so vehicles spread over small streets as well as main ones (see \cite{Uruena2017} for details)}. \textcolor{black}{All vehicles send CAM traffic according to ETSI rules~\cite{etsiCA}.}

We work with two scenarios: 1) we disseminate an emergency message to the largest Destination Area allowed (i.e., 10~km$^2$), and 2) we disseminate emergency messages in a smaller area, which could be interesting for different applications in urban scenarios and where we have compared the multi-hop broadcast mechanisms with \ac{SHB}. 

\begin{table*}[tbh!]
\caption{Average number of vehicles inside the Destination Area \textcolor{black}{(maximum destination area)}}
\centering
\begin{tabularx}{0.3\textwidth}{| >{\centering\arraybackslash}X | >{\centering\arraybackslash}X | }
\hline
\textbf{Veh. in scenario} & \textbf{Avg. veh. in dest. area}\\\hline
 600 &  586 \\ \hline
1200 & 1025 \\ \hline
1800 & 1555 \\ \hline
2400 & 2161 \\ \hline
\end{tabularx}
\label{table:urban_veh_roi}
\end{table*}

The first scenario consists of a stationary vehicle in the center of the map, located at the intersection of one major avenue and a medium-sized, four-lane street. The vehicle starts sending \acp{DENM} to a circular Destination Area with a radius of 1.784~km. 30 DEMN messages are generated at a 1~Hz frequency and they are configured as multi-hop GeoBroadcast messages (i.e., to be forwarded). For the first scenario, in terms of vehicle density, we simulate four different vehicle quantities: 600, 1200, 1800, and 2400 vehicles. Table~\ref{table:urban_veh_roi} shows the average quantity of vehicles in total and within the Destination Area for every density.

The second scenario uses a medium number of vehicles, i.e., a density of 1200 vehicles. We select four other locations around the map in order to have, additionally to the original point at the center of the map, source points in the southeast, southwest, northeast, and northwest of the map. These points are located on small to medium-sized streets, either at intersections or in the middle of the street. The variety in the types of streets allows for a wide range of situations, from square blocks in the southeast, to alley-like streets in the northwest. The radius of the Destination Area, which applies for multi-hop GeoBroadcast is 500\,m (i.e. 0.78~km$^2$), while \ac{SHB} does not specify a destination area and it reaches receivers as far as the radio channel allows.

We use the same simulation tools as in section~\ref{sec:highway}, with the addition of urban-related configurations. The impact of obstacles (e.g., buildings) in signal propagation has been modeled with two attenuation models: the Two-Ray Interference Model \cite{Sommer:2012}, and the Obstacle Shadowing model \cite{Sommer:2011}. The parameters of the obstacle model are shown in Table~\ref{tbl:simpars}.

\subsection{Results}
The performance metrics are the same used in the highway scenario: number of transmissions, \acf{PDR}, and end-to-end delay.
\label{subsec:results_urban}
\subsubsection{Maximum Destination Area}
\label{subsubsec:results_urban_max}

\begin{figure}[tbh!]
    \centering
    \includegraphics[trim={4cm 0 4cm 0},clip,width=\textwidth]{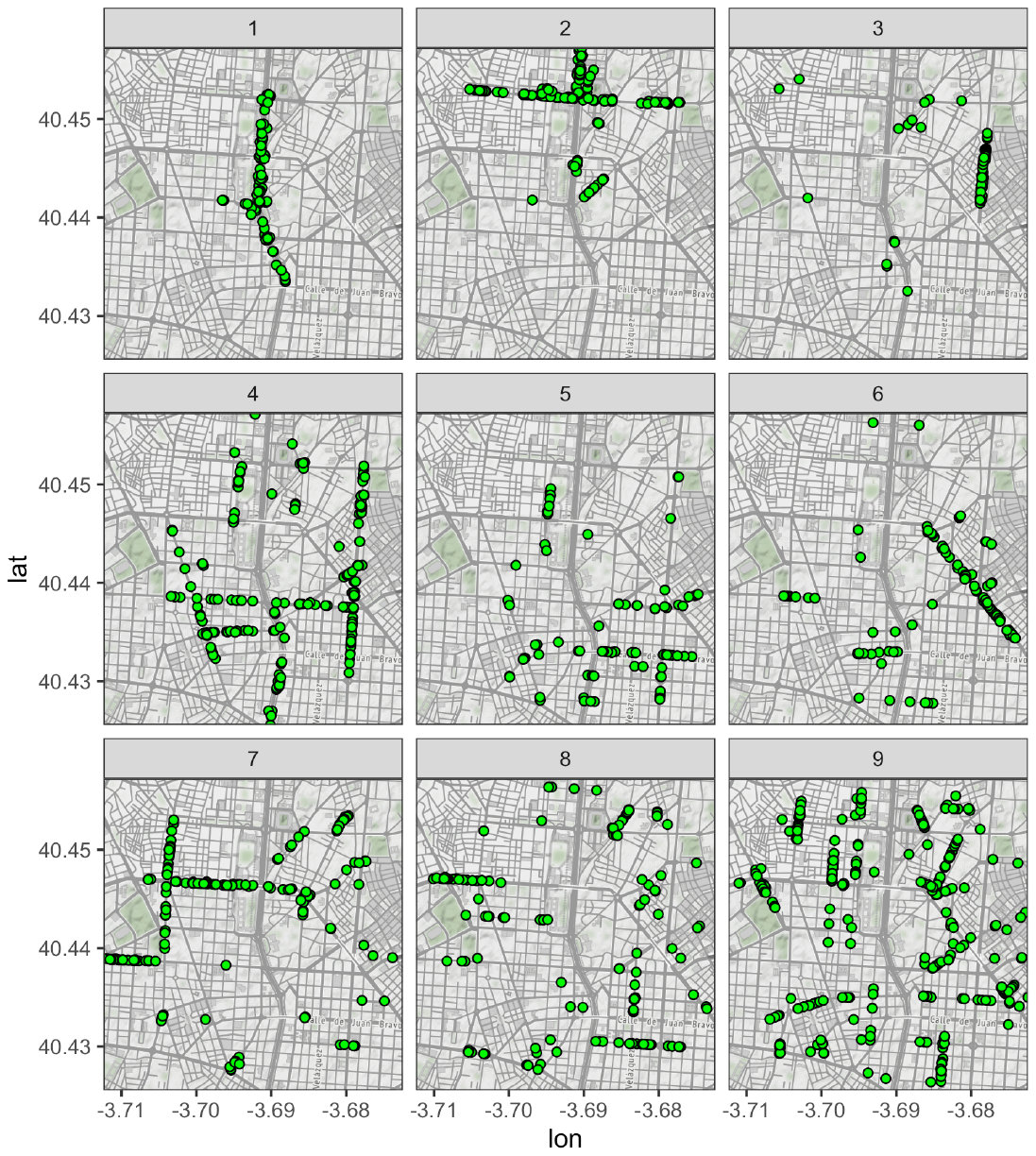}
    \caption{Message dissemination for ETSI CBF in an urban scenario \textcolor{black}{(maximum destination area)}}
    \label{fig:hops_urban}
\end{figure}

Fig.~\ref{fig:hops_urban} shows how the first \ac{DENM} sent is disseminated using \ac{ETSI} \ac{CBF}. Each green dot represents the first reception and buffering of a message, and every frame represents the hop count in which the message was received. The first hop disseminates the message from the source along the main avenue where the originating vehicle is located, with only a few receptions occurring on side roads. From there, the message is forwarded by other vehicles (our simulations do not include any \acp{RSU}), and we can see several phenomena occurring: 1) the presence of obstacles causes messages to get disseminated along roads, with vehicles in intersections serving as relays to allow messages to change directions (e.g., the clusters on the north of hop 2 and the west of hop 7); and 2) information is not disseminated radially away from the center, i.e., a message can "go away from" and "come closer to" the center of the Destination Area with every hop. An example of this can be seen when comparing hop 1 and hop 7, where  \textcolor{black}{vehicles} closer to the source receive the message before  \textcolor{black}{vehicles} farther away (e.g., in hop 7, we even see receptions outside of the Destination Area). 

\begin{figure}[tbh!]
    \centering
    \begin{subfigure}[b]{0.45\textwidth}
      \centering
      \caption{Transmissions for ETSI CBF}
      \includegraphics[width=1.2\textwidth]{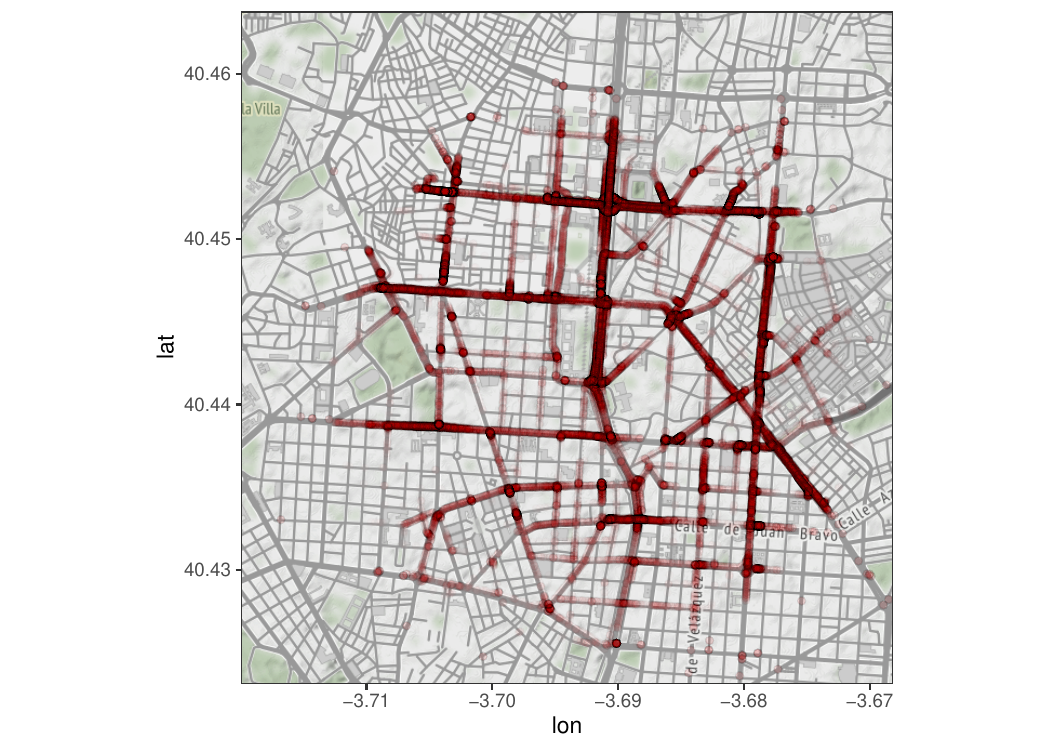}
      \label{fig:tx_cbf_map}
    \end{subfigure}
    \begin{subfigure}[b]{0.45\textwidth}
      \centering
      \caption{Transmissions for S-FoT+}
      \includegraphics[width=1.2\textwidth]{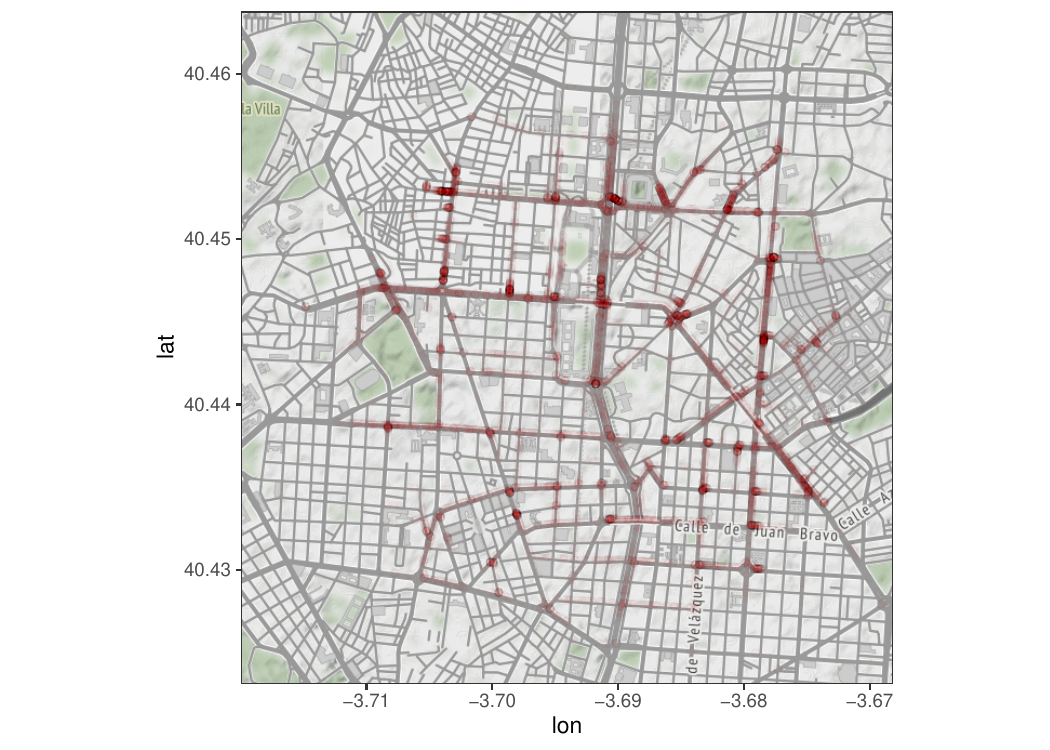}
      \label{fig:tx_sfotp_map}
    \end{subfigure}
    \begin{subfigure}[b]{0.45\textwidth}
      \centering
      \caption{DENM receptions for ETSI CBF}
      \includegraphics[width=1.2\textwidth]{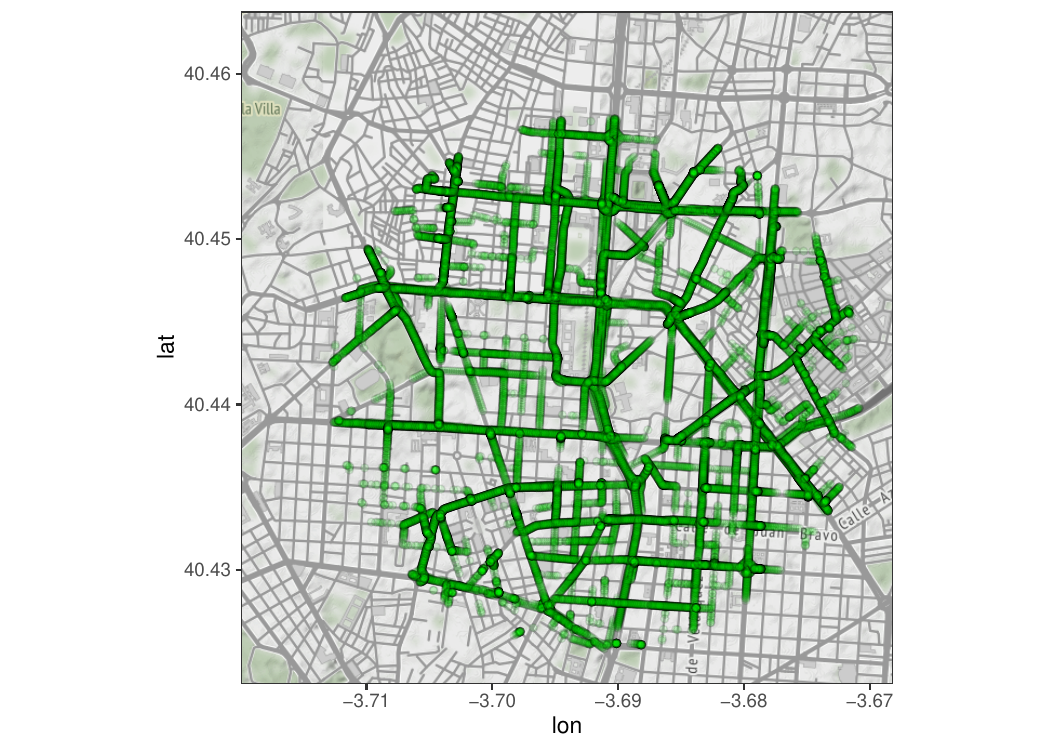}
      \label{fig:rx_cbf_map}
    \end{subfigure}
    \begin{subfigure}[b]{0.45\textwidth}
      \centering
      \caption{DENM receptions for S-FoT+}
      \includegraphics[width=1.2\textwidth]{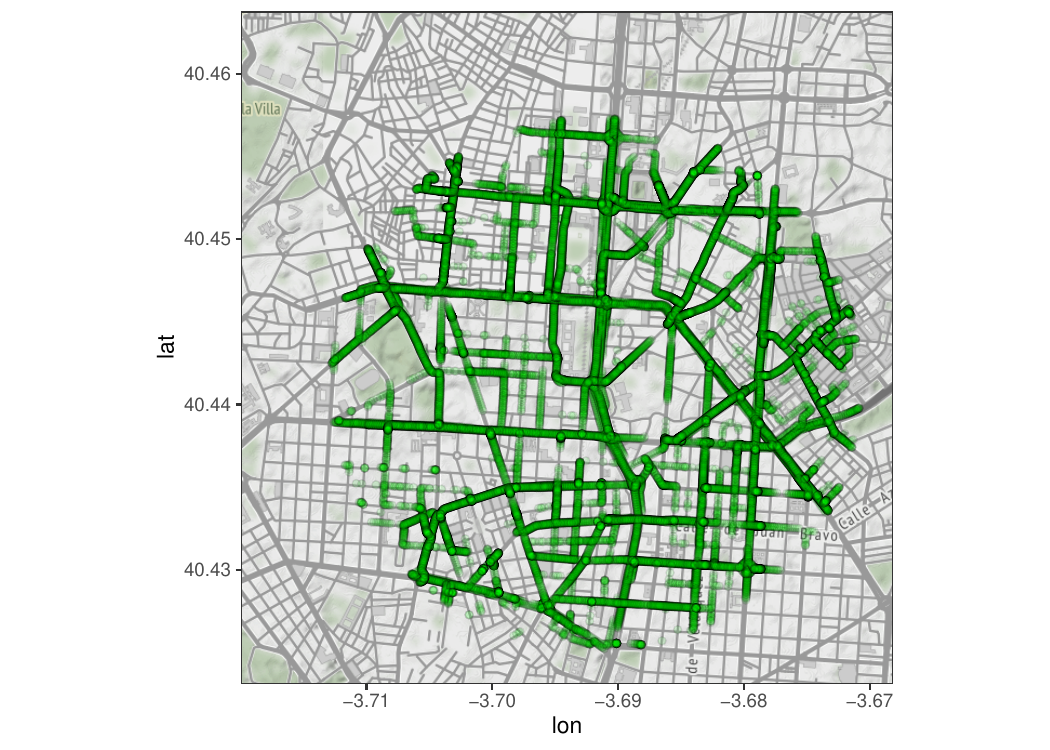}
      \label{fig:rx_sfotp_map}
    \end{subfigure}
    \caption{Transmissions and receptions footprint for the 2400 veh. scenario \textcolor{black}{(maximum destination area)}}
    \label{fig:tx_rx_maps}
\end{figure}

Fig.~\ref{fig:tx_rx_maps} shows the effect of all the mechanisms included in S-FoT+ on the overall efficiency of GeoBroadcasting in the highest simulated density. The bottom of the figure shows the coordinates of effective \ac{DENM} receptions at the facilities layer, i.e., only the first time the message is consumed, given that \ac{ETSI} \ac{CBF} is prone to send duplicate messages up to the \ac{DEN} basic service. Both \ac{ETSI} \ac{CBF} and S-FoT+ show a very similar reception density, but the transmission footprint -- shown at the top of the figure -- is significantly different. The footprint for the \ac{ETSI} \ac{CBF} mechanism is not only denser but also some of the dots that represent transmissions are located outside of the Destination Area, as it is evident towards the outer part of the circle. 

\begin{figure}[tbh!]
    \centering
    \includegraphics[width=0.8\textwidth]{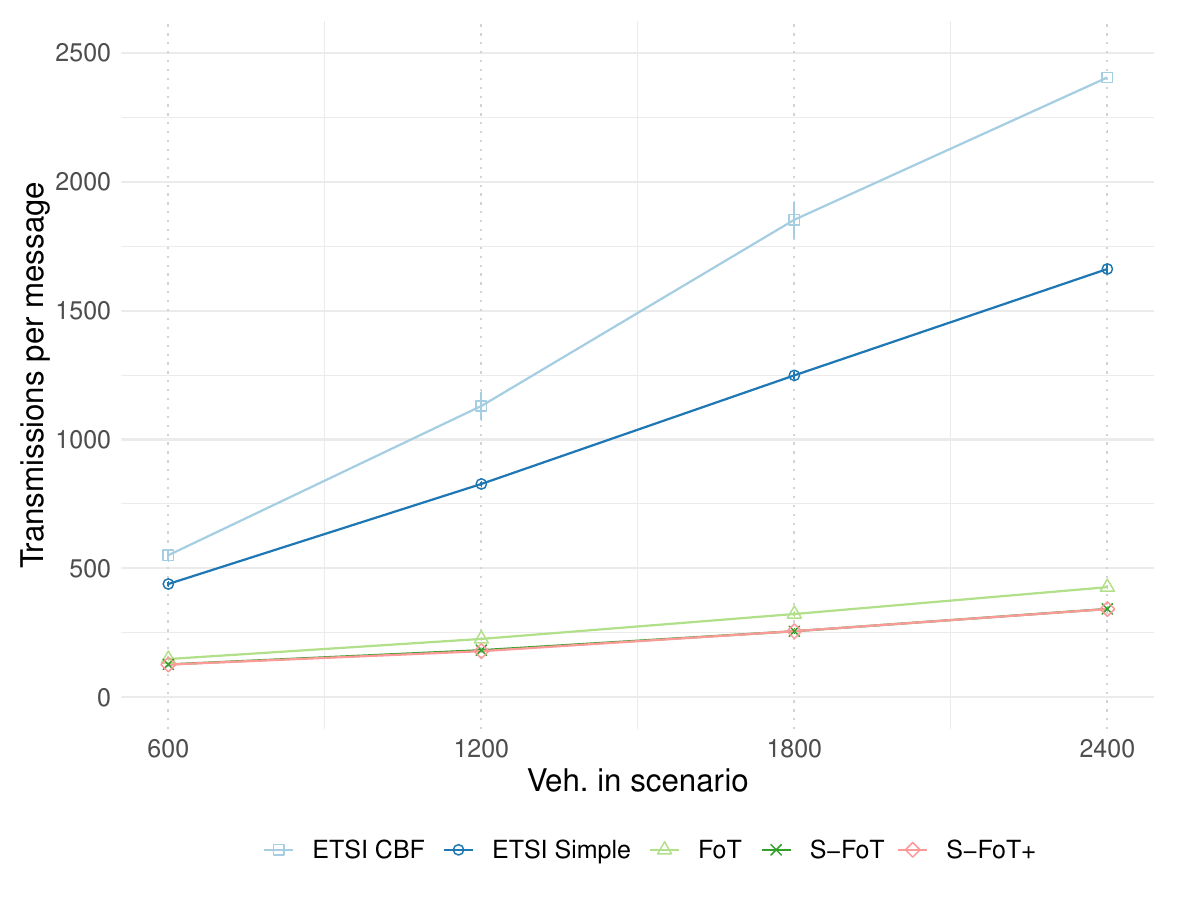}
    \caption{Transmissions per density in the urban scenario \textcolor{black}{(maximum destination area)}}
    \label{fig:tx_urban}
\end{figure}

Fig.~\ref{fig:tx_urban} and Table~\ref{table:urban_txr} show that, for the urban scenario, the quantity of transmissions grows with the number of vehicles for all mechanisms. However, in line with the results from section~\ref{sec:highway}, \ac{FoT}-based mechanisms use only a fraction of the messages standard \ac{ETSI} mechanisms use. The difference between \ac{FoT} and the slotted variants is lower than in highway scenarios, yet S-FoT+ uses between 14\% and 21\% fewer transmissions than \ac{FoT} to achieve similar results. Nevertheless, this smaller difference can be attributed to the fact that extremely-high medium congestion is not a common occurrence in urban scenarios, and also transmissions reach shorter distances than in highways, where there are fewer obstacles.

\begin{table*}[tbh!]
\caption{Average number of transmissions \textcolor{black}{per message} for the urban scenario \textcolor{black}{(maximum destination area)}}
\centering
\begin{tabularx}{\textwidth}{| >{\raggedright\arraybackslash}X |  >{\raggedleft\arraybackslash}X | >{\raggedleft\arraybackslash}X | >{\raggedleft\arraybackslash}X | >{\raggedleft\arraybackslash}X | >{\raggedleft\arraybackslash}X |}
\hline
\textbf{Veh. in scenario} & \textbf{ETSI CBF} & \textbf{ETSI Simple} & \textbf{FoT} & \textbf{S-FoT} &\textbf{S-FoT+} \\\hline
 600 & 550.80 & 439.15 &  147.92 &  126.97 &  126.43\\ \hline
1200 & 1,129.96 & 827.32 &  226.08 &  182.23 &  178.42\\ \hline
1800 & 1,852.71 & 1,248.78 &  322.71 &  255.88 &  256.05\\ \hline
2400 & 2,404.84 & 1,661.72 & 426.89 & 342.39 & 341.88\\ \hline
\end{tabularx}
\label{table:urban_txr}
\end{table*}

In terms of \ac{PDR} (see Table~\ref{table:urban_pdr}), there is an effect of vehicle sparsity in combination with signal propagation phenomena: \ac{PDR} decreases for all mechanisms when fewer vehicles are present since there are fewer opportunities to broadcast a message when vehicles cannot \textit{see} each other (i.e., due to lack of line of sight or because there are no vehicles within one hop of each other). However, more conservative mechanisms (including \ac{ETSI} Simple GeoBroadcast) do achieve a higher \ac{PDR} when compared to ETSI CBF.

\begin{table*}[tbh!]
\caption{Average Packet-Delivery Ratio for the urban scenario \textcolor{black}{(maximum destination area)}}
\centering
\begin{tabularx}{\textwidth}{| >{\raggedright\arraybackslash}X |  >{\raggedleft\arraybackslash}X |
>{\raggedleft\arraybackslash}X | >{\raggedleft\arraybackslash}X | >{\raggedleft\arraybackslash}X | >{\raggedleft\arraybackslash}X |}
\hline
\textbf{Veh. in scenario} & \textbf{ETSI CBF} & \textbf{ETSI Simple} & \textbf{FoT} & \textbf{S-FoT} &\textbf{S-FoT+} \\\hline
 600 & 0.7674  & 0.8498  & 0.8685  & 0.8508  & 0.8519 \\ \hline
1200 & 0.8808  & 0.9102  & 0.9235  & 0.9084  & 0.9140 \\ \hline
1800 & 0.9299  & 0.9379  & 0.9488  & 0.9478  & 0.9478 \\ \hline
2400 & 0.9157  & 0.9419  & 0.9506  & 0.9497  & 0.9514 \\ \hline
\end{tabularx}
\label{table:urban_pdr}
\end{table*}

\begin{figure}[tbh!]
    \centering
    \includegraphics[width=0.8\textwidth]{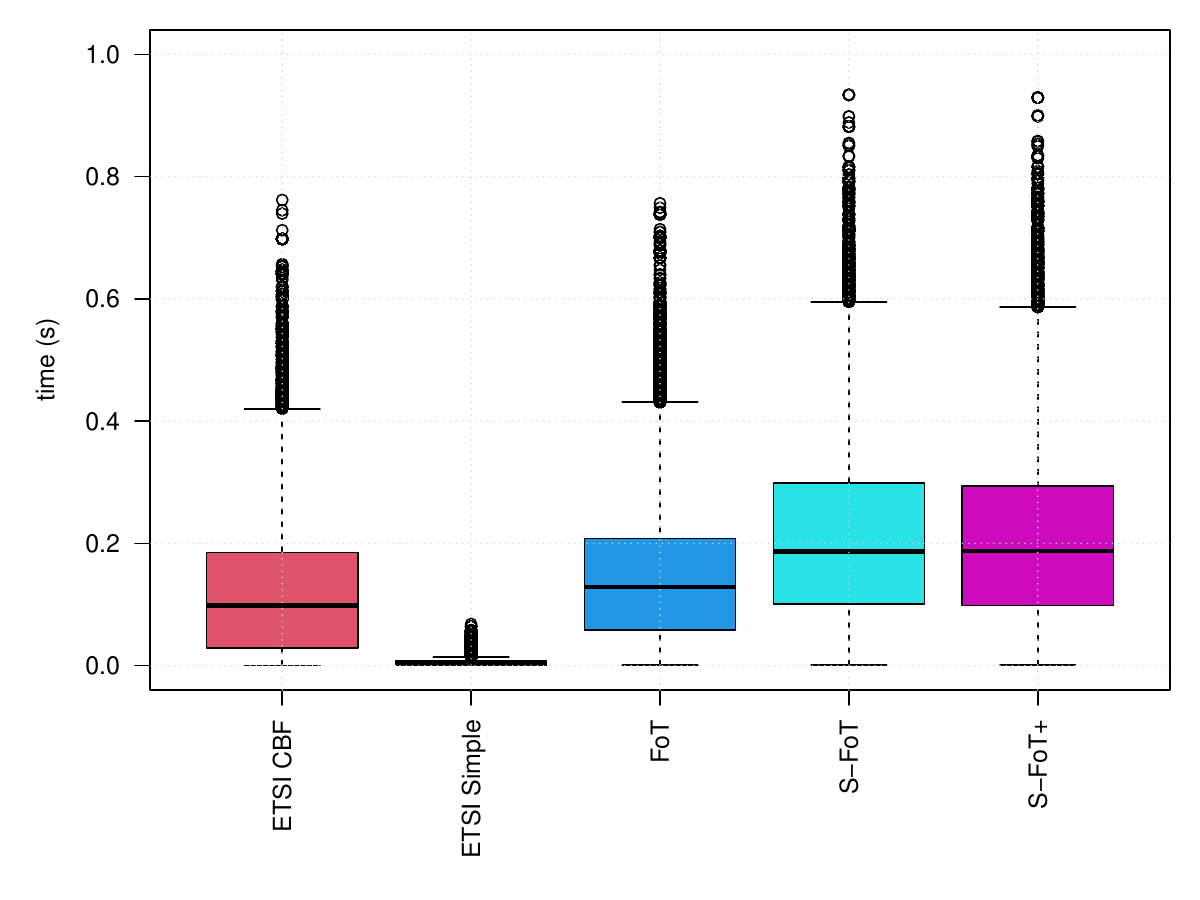}
    \caption{\textcolor{black}{End-to-end delay for the 600-vehicle urban scenario \textcolor{black}{(maximum destination area)}}}
    \label{fig:e2e_urban}
\end{figure}

\textcolor{black}{Fig.~\ref{fig:e2e_urban} shows the end-to-end delay for all mechanisms in the urban scenario with 600 vehicles. There is an effect of Slotted CBF that is visible in S-FoT and S-FoT+ when compared to the CBF mechanisms without slots (i.e., ETSI CBF and FoT). Nevertheless, even extreme outlier values stay below 1\,s, and the maximum for slotted mechanisms are around 600\,ms, which added to the PDR values, shows that S-FoT and S-FoT+ manage to cover most vehicles in the area within that period.}

The main takeaway from this evaluation is that FoT-based forwarding, and especially those using slots, are equally effective and efficient in urban and highway scenarios, which confirms the versatility of these mechanisms. For example, none of our simulation scenarios consider the use of infrastructure (e.g. \acp{RSU}), which means that total reachability for warning messages can be achieved by fully decentralized implementations, such as ETSI ITS-G5.

\subsubsection{Comparison of Multi-hop Broadcasting to Single-hop Broadcasting in small areas}

Section~\ref{subsubsec:results_urban_max} shows that there are different phenomena affecting message propagation in urban environments. Fig.~\ref{fig:hops_urban} and Table~\ref{table:urban_pdr} show that: 1) messages are relayed by other \textcolor{black}{vehicles} in order to surmount obstacles and go around blocks, and 2) even with the help of relays, not all \textcolor{black}{vehicles} in the Destination Area receive all messages. Furthermore, considering the overhead cost of multi-hop mechanisms, another question arises regarding the appropriateness of the multi-hop schemes if applications only require reaching shorter distances. In other words, while it is straightforward that multi-hop is necessary to cover distances and areas such as those in the highway and urban scenarios explored in Sec.~\ref{subsec:results_highway} and \ref{subsubsec:results_urban_max}, if an application requires to cover a shorter distance or reach a smaller area, i.e., within one hop, \ac{SHB} becomes a viable option.

We evaluate the performance of multi-hop mechanisms (ETSI \ac{CBF}, ETSI Simple GeoBroadcast, FoT, and S-FoT+) in reduced areas, and compare it with a benchmark given by ETSI \ac{SHB}. We selected five locations on the map where events are notified using \acp{DENM}. The number of vehicles in the scenario is 1200. The events occur in streets and avenues with different features, and are located in intersections and in between streets. For the multi-hop mechanisms, a Destination Area is defined: a circle with a radius of 500\,m. ETSI \ac{SHB} does not require the definition of a \textit{Destination Area}, and messages are received as far as the radio medium allows, which as per Fig.\ref{fig:pdr_shb_highway}, can surpass 1000\,m. For the comparison of the different mechanisms, we consider the \ac{PDR} only within the Destination Area.

\begin{figure}[tbh!]
    \centering
    \begin{subfigure}[b]{0.7\textwidth}
      \centering
      \caption{Receptions for ETSI SHB}
      \includegraphics[width=\textwidth]{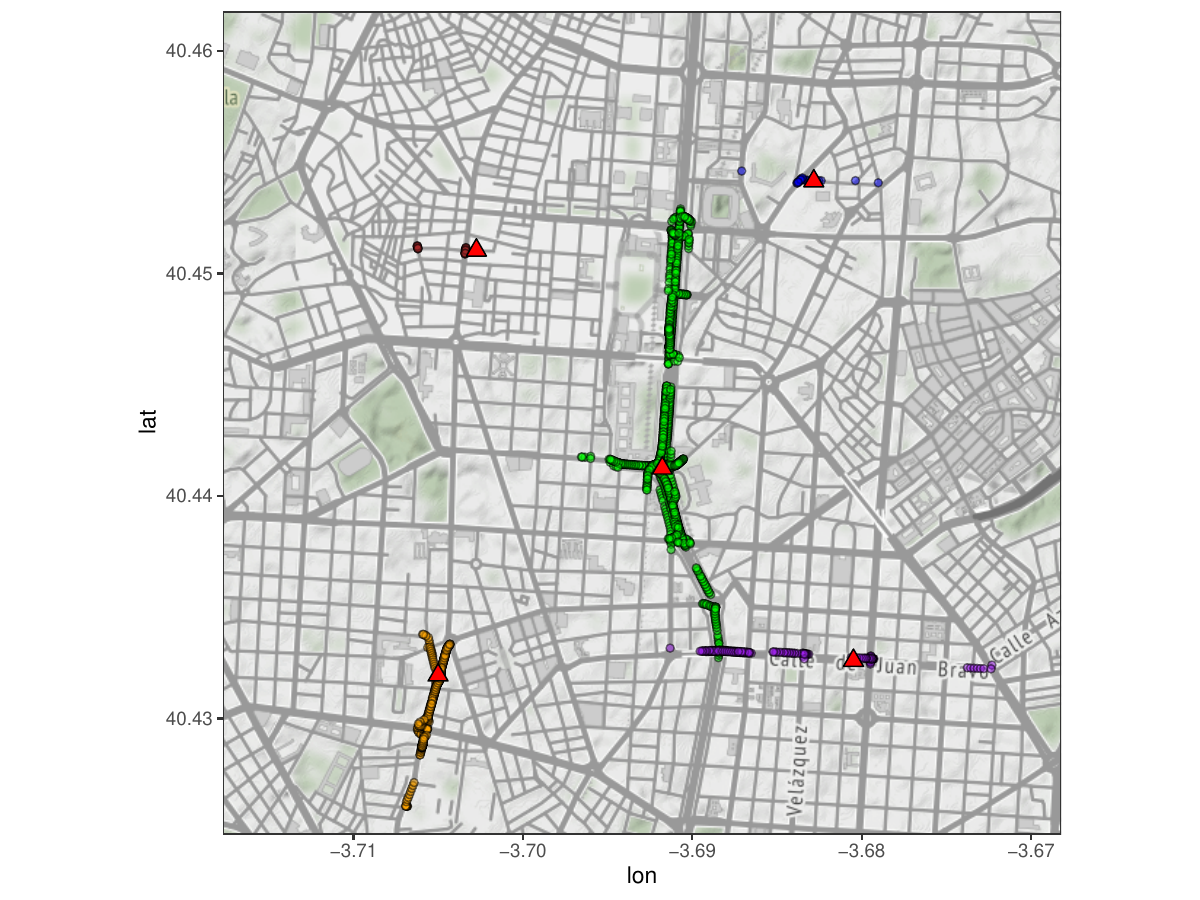}
      \label{fig:rx_shb_5p}
    \end{subfigure}
    \begin{subfigure}[b]{0.7\textwidth}
      \centering
      \caption{Receptions for S-FoT+}
      \includegraphics[width=\textwidth]{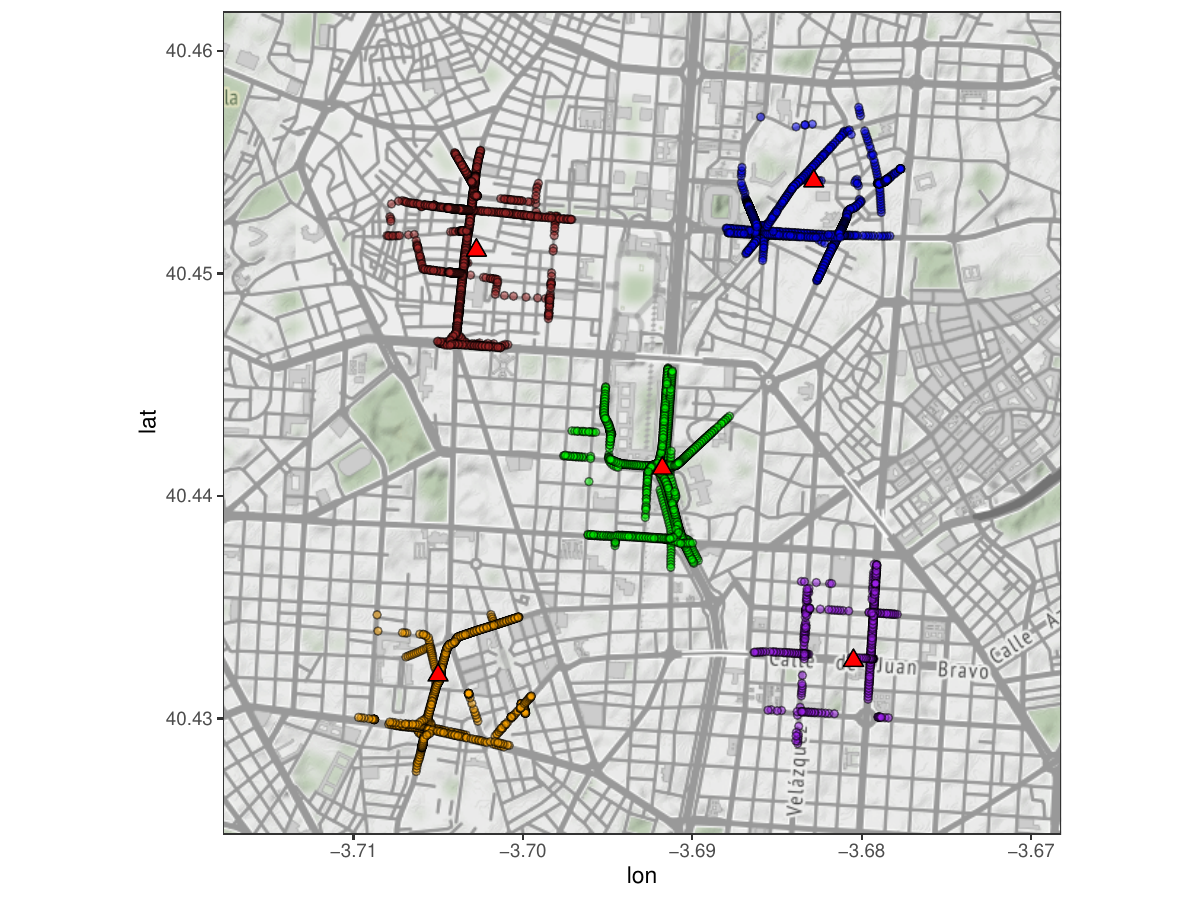}
      \label{fig:rx_sfotp_5p}
    \end{subfigure}
    \caption{Comparison of receptions in five events for Single-hop Broadcast and S-FoT+ \textcolor{black}{(with 1200 vehicles and small destination area)}}
    \label{fig:rx_5p}
\end{figure}

Fig.~\ref{fig:rx_5p} shows the receptions obtained by ETSI \ac{SHB} and S-FoT+ for the five locations. Each event is represented by a different color: green for the source at the center of the map, orange for the southwest, purple for the southeast, blue for the northeast, and brown for the northwest. The source \textcolor{black}{vehicles} are represented by red triangles. Starting from the overall picture, one result is evident: the multi-hop capability of S-FoT+ enables it to disseminate messages along more axes than \ac{SHB}. The north sides of Fig.\ref{fig:rx_shb_5p} and Fig.\ref{fig:rx_sfotp_5p} show that, unless multi-hop is in place, events will not be received but by those vehicles in the line-of-sight of the source  \textcolor{black}{vehicle}.

The center of the map in Fig.\ref{fig:rx_shb_5p}, however, shows that \ac{SHB} reaches vehicles farther along the main road where the source  \textcolor{black}{vehicle} is located. This does not mean that these messages are not received by vehicles in the multi-hop GeoBroadcast scenario, but they are discarded since the vehicles are outside the Destination Area. Nevertheless, the message is not well propagated sideways -- the street that branches northwest from the center is not well covered by \ac{SHB}. On the other hand, S-FoT+ shows uniform coverage not only along those streets but also in the blocks around the source \textcolor{black}{vehicle}, and the message is relayed around corners. Furthermore, while receptions occur at longer distances in the \ac{SHB} scenario,  \textcolor{black}{vehicles} farther away from the source also suffer from losses stemming from radio propagation phenomena.

\begin{figure}[tbh!]
    \centering
    \includegraphics[width=0.8\textwidth]{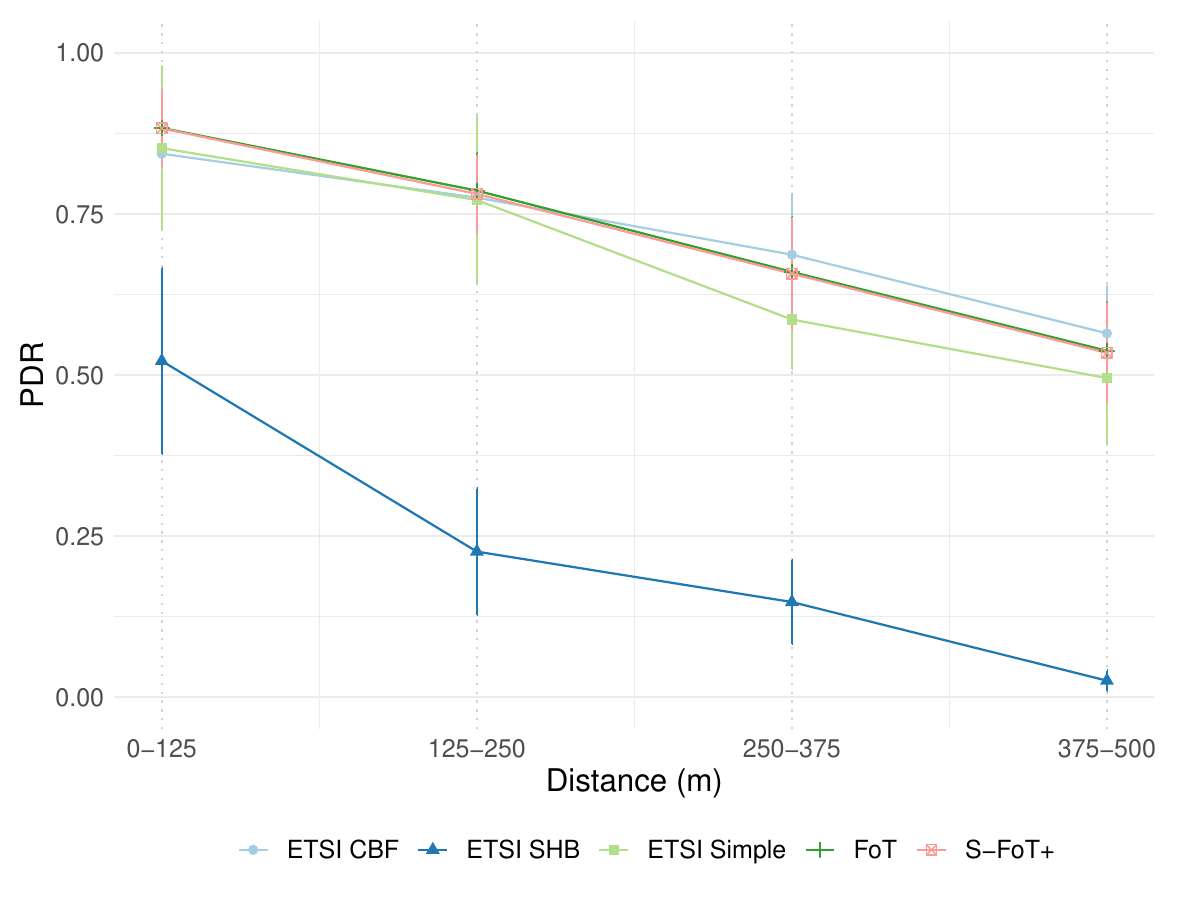}
    \caption{Average Packet-Delivery Ratio over distance for urban events \textcolor{black}{(with 1200 vehicles and small destination area)}}
    \label{fig:pdr_5p}
\end{figure}

\textcolor{black}{Fig.~\ref{fig:pdr_5p} shows the effect of radio propagation on \ac{SHB} in an urban environment and the gains in reachability that come with the use of multi-hop schemes. Even at short distances (0 to 125\,m), \ac{SHB} only manages to deliver packets to an average of 52\% of neighboring vehicles. Table~\ref{table:tx_5p} shows that there is over an order of magnitude in overhead between \ac{SHB} and S-FoT+, but it comes in exchange for a significant difference in performance: the highest average PDR for \ac{SHB} (52\% at 0-125\,m) is lower than the worst average results for S-FoT+ (53\% at 375-500\,m). Therefore, one important conclusion from this evaluation is that, in urban environments, multi-hop schemes are needed when applications require reliable coverage, even over short distances or small areas.}

\begin{table*}[tbh!]
\caption{Average number of transmissions \textcolor{black}{per message for urban events }\textcolor{black}{(with 1200 vehicles and small destination area)}}
\centering
\begin{tabularx}{\textwidth}{| >{\centering\arraybackslash}X | >{\centering\arraybackslash}X |  >{\centering\arraybackslash}X | >{\centering\arraybackslash}X | >{\centering\arraybackslash}X | >{\centering\arraybackslash}X | }
\hline
& ETSI CBF & ETSI Simple & FoT & S-FoT+ & ETSI SHB\\\hline
Avg. Tx. & 91.86 &  59.26 & 12.86 & 12.90 & 1.00 \\ \hline
\end{tabularx}
\label{table:tx_5p}
\end{table*}

\section{Conclusions}
\label{sec:conclusions-fw}

This paper has reviewed the improvements proposed in the literature for \ac{ETSI} \ac{ITS} \acf{CBF} and evaluated several of them, as well as \ac{ETSI} Simple GeoBroadcast forwarding, in highway and urban scenarios. Quite surprisingly, \ac{ETSI} Simple GeoBroadcast forwarding has a better performance than \ac{ETSI} \ac{CBF} in all cases, due to the lack of \ac{DPD} in the standard \ac{ETSI} \ac{CBF} algorithm. Nevertheless, this paper also shows that the performance of the \ac{CBF} algorithm can be greatly improved by implementing the mechanisms proposed in \cite{Amador2022}. The result is a significant decrease in the total number of transmissions compared to both the original \ac{ETSI} \ac{CBF} and Simple GeoBroadcast forwarding algorithms, while keeping a high \ac{PDR}. In particular, this result is maintained in urban scenarios, where the proposed changes to the \ac{CBF} forwarding algorithm, \textcolor{black}{using a} reduced number of transmissions, \textcolor{black}{allow messages} to reach \textcolor{black}{most} vehicles in the Destination Area.

Moreover, this paper has introduced two further improvements to \ac{ETSI} \ac{CBF}: 1) \acf{S-CBF}, which effectively prevents collisions when packets are received by several vehicles beyond $DIST_{MAX}$ (1~km); and 2) FoT+, a further improvement to our previous \ac{FoT} mechanism that guarantees that a forwarded \ac{DENM} waits at the \ac{CBF} buffer (and thus can be cancelled by a retransmission) instead of at a \ac{DCC} queue, even when higher-priority \acp{CAM} are sent. Therefore, we can conclude that S-FoT+, which includes all improvements in \cite{Amador2022} as well as S-CBF and FoT+, has a similar or better performance than \ac{ETSI} \ac{CBF} and Simple GeoBroadcast forwarding in both \ac{PDR} and end-to-end delay, while requiring significantly fewer transmissions, in all the evaluated scenarios and vehicle densities.

Finally, we demonstrated that \ac{SHB} exhibits significant shortcomings in urban settings. Even in small Destination Areas, the effect of obstacles (i.e., city blocks) and the intrinsic inability of the protocol to go around corners deems it unsuitable in many scenarios that are common in cities. We have shown that multi-hop GeoBroadcast has a better performance in terms of \ac{PDR}, but it comes at the cost of almost two orders of magnitude more transmissions for the case of \ac{ETSI}~\ac{CBF}. However, this cost is reduced when using S-FoT+, which is an interesting trade-off to consider for certain application requirements.

\bibliography{mybibfile}

\end{document}